\documentclass[12pt]{iopart}
\usepackage{iopams} 
\pdfoutput=1
\usepackage[utf8]{inputenc}
\usepackage{cite}
\usepackage{graphicx}
\usepackage{upgreek}
\usepackage[english]{babel}
\usepackage{units}
\usepackage{booktabs}

\graphicspath{{pics/}}

\begin{document}

\title[Advanced fluid modelling and PIC/MCC simulations of low-pressure ccrf discharges]{Advanced fluid modelling and PIC/MCC simulations of low-pressure ccrf discharges}

\author{M. M. Becker$^{1}$, H. Kählert$^2$, A. Sun$^{1,3}$, M. Bonitz$^2$, and D.~Loffhagen$^1$}

\address{$^1$Leibniz Institute for Plasma Science and Technology, Felix-Hausdorff-Str. 2, 17489 Greifswald, Germany}
\address{$^2$Institut für Theoretische Physik und Astrophysik, Christian-Albrechts-Universität zu Kiel, Leibnizstr. 15, 24098 Kiel, Germany}
\address{$^3$State Key Laboratory of Electrical Insulation and Power Equipment, School of Electrical Engineering, Xi'an Jiaotong University, 710049 Xi'an, People's Republic of China}

\ead{markus.becker@inp-greifswald.de}
\vspace{10pt}

\begin{abstract}

Comparative studies of capacitively coupled radio-frequency discharges in helium and argon at pressures between 10 and 80 Pa  are presented applying two different fluid modelling approaches as well as two independently developed particle-in-cell/Monte Carlo collision (PIC/MCC) codes. 
The focus is on the analysis of the range of applicability of a recently proposed fluid model including an improved drift-diffusion approximation for the electron component as well as its comparison with fluid modelling results using the classical drift-diffusion approximation and benchmark results obtained by PIC/MCC simulations. Main features of this time- and space-dependent fluid model are given. 
It is found that the novel approach shows generally quite good agreement with the macroscopic properties derived by the kinetic simulations and is largely able to characterize qualitatively and quantitatively the discharge behaviour even at conditions when the classical fluid modelling approach fails. 
Furthermore, the excellent agreement between the two PIC/MCC simulation codes using the velocity Verlet
method for the integration of the equations of motion
%
verifies their accuracy and applicability.

\end{abstract}

%
%
%
%
%

\section{Introduction}

Capacitively coupled radio-frequency (ccrf) discharge plasmas are widely used in plasma processing technologies. Typical examples for the application of ccrf discharges are plasma-enhanced chemical vapor deposition and plasma etching~\cite{LiebermannBook}. Besides experimental diagnostics, numerical modelling and simulation of ccrf discharges provide established tools to get detailed insights into the discharge physics~\cite{vanDijk-2009-ID2560,Donko-2012-ID3151,Alves-2012-ID3406}. 

Particle-in-cell/Monte Carlo collision (PIC/MCC) simulation is the most recognised method for the theoretical description of ccrf discharges~\cite{Donko-2012-ID3151,Vahedi-1993-ID3616,Matyash-2007-ID3787,Bronold-2007-ID2423,Turner-2013-ID3004,Erden-2014-ID3505,Eremin-2016-ID3872,Sun-2016-ID3996}. In the PIC/MCC method, a collection of particles is followed in space and time taking into account particle-particle and particle-wall interactions as well as the effect of the self-consistently determined space charge field~\cite{vanDijk-2009-ID2560,Donko-2011-ID2668}.
Fluid (or continuum) models, which are based on a hydrodynamic description of the plasma, provide an alternative approach for the analysis of ccrf discharges~\cite{Barnes-1987-ID544,Passchier-1993-ID3052,Boeuf-1995-ID3041,Hammond-2002-ID3321,Salabas-2002-ID1762,Alves-2012-ID3406}. 
Compared to PIC/MCC simulations, the numerical solution of fluid models is computationally less demanding and hence more attractive for practical applications. A disadvantage of the fluid approach is its limited application range when low-pressure ccrf discharges are considered.
Here, the fluid description is applicable as long as the charged particles' mean free path is much smaller than the characteristic dimension of the discharge~\cite{Belenguer-1990-ID2897}.
%
%
Different modelling approaches have been developed that aim at a combination of the advantages of kinetic simulations and fluid models in so-called hybrid methods~\cite{Bogaerts-1999-ID3030,Donko-2006-ID2349,Li-2008-ID3661,Eylenceoglu-2015-ID3713,Eremin-2016-ID3828} or at an improvement of the accuracy of the classical fluid description~\cite{Belenguer-1990-ID2897,Robson-2005-ID2280,Nicoletopoulos-2012-ID2717,Robson-2012-ID2932,Rafatov-2012-ID2864,Chen-2004-ID2132,Dujko-2013-ID3242,Becker-2013-ID3118,Becker-2013-ID2934}.
Since the latter is in the focus of the present paper, an overview of recent approaches is given here.

In non-thermal plasmas, the energy is mainly delivered through the electrons. Therefore, an accurate description of the electron component is crucial for the reliability of an integral plasma model~\cite{Derzsi-2009-ID2552} and hence researchers have spent large effort to improve the description of the electron component. 
Earlier works aimed at a customized description of fast electrons by means of so-called beam models, see, e.g.,~\cite{Belenguer-1990-ID2897,Kulikovsky-1991-ID3056,Capriati-1993-ID3047} and references therein.
Robson \textit{et al.}~\cite{Robson-2005-ID2280} have introduced a physically based fluid model for electrons in low-temperature plasmas based on the so-called \textit{heat flux ansatz}.
This approach has been applied in Ref.~\cite{Nicoletopoulos-2008-ID2478} to describe  periodic electron structures in a constant electric field by means of a fluid model and it
has been generalised in Ref.~\cite{Nicoletopoulos-2012-ID2717} to electron swarms under the influence of nonuniform electric fields.
In another work of Robson \textit{et al.}~\cite{Robson-2012-ID2932}, the accuracy of fluid models for light particles has been improved by a direct substitution technique that uses swarm transport data instead of cross sections for the calculation of the collision terms.
Rafatov~\textit{et al.}~\cite{Rafatov-2012-ID2864} have proposed a fluid model that includes a nonlocal ionization source term in order to overcome fundamental shortcomings of the classical fluid description.
A non-local collisionless electron heat flux has been considered in Ref.~\cite{Chen-2004-ID2132} to get an improved description of the collisionless electron heating effect in low-pressure, high-frequency ccrf discharges.
Furthermore, fluid models for electrons in non-thermal plasmas based on a four-moment description have been developed~\cite{Dujko-2013-ID3242,Becker-2013-ID3118}. 
Note that in the frame of the hydrodynamic plasma description two moment equations (particle density and flux) are commonly considered for heavy particles, while three moment equations (particle density, flux and mean energy density) are usually taken into account for the electron component~\cite{Kim-2005-ID2245,Alves-2012-ID3406}.
In Refs.~\cite{Markosyan-2013-ID3243,Markosyan-2015-ID3809} it has been shown that the four-moment model proposed in Ref.~\cite{Dujko-2013-ID3242} is more accurate than conventional fluid approaches for the theoretical description of negative streamer fronts in nitrogen and neon, while the classical fluid model using the local field approximation (LFA)~\cite{Ward-1962-ID93,Boeuf-1987-ID545} gives reasonably good results under these conditions, too~\cite{Markosyan-2015-ID3809}.
In contrast, Grubert \textit{et al.}~\cite{Grubert-2009-ID2551} have shown that the LFA is not applicable for the investigation of low-pressure gas discharge plasmas and the local mean energy approximation (LMEA)~\cite{Boeuf-1995-ID3041}
%
should be used instead.
A novel LMEA based drift-diffusion approximation for electrons has been derived in Ref.~\cite{Becker-2013-ID2934} from the four-moment model proposed in Ref.~\cite{Becker-2013-ID3118}. This model has been found to be more accurate than the classical LMEA based fluid description for dc glow discharge plasmas at low and atmospheric pressure~\cite{Becker-2013-ID2934,Becker_dissbook}.
The main advantage of this drift-diffusion approach established in Ref.~\cite{Becker-2013-ID2934} compared to most of the other modified fluid descriptions is that it is not limited to specific discharge conditions and it does not increase the computational cost~\cite{Becker_dissbook}.


Comparisons between fluid and particle methods for the simulation of low-pressure ccrf discharges have been carried out before, e.g., in Refs.~\cite{Surendra-1993-ID812,Lymberopoulos-1995-ID3301,Diomede-2008-ID4004,Surendra-1995-ID3040,Turner-2013-ID3004}. 
From the rigorous comparison of different PIC/MCC simulation and fluid modelling results for helium ccrf discharges in the pressure range from 4 to 40\,Pa in Surendra~\cite{Surendra-1995-ID3040} it can be concluded that the degree of agreement between fluid and PIC/MCC simulation methods strongly depends on the gas pressure and is also divergent for different plasma properties.
For the lowest pressure considered in~\cite{Surendra-1995-ID3040}, larger differences between PIC/MCC and fluid results of about 50 to 60\,\% have been found for the predicted rf voltage and plasma density while generally good agreement for other important properties, such as the ion flux to the electrodes (error of 20 to 30\,\%), has been reported. 
For the same benchmark situation as in Ref.~\cite{Surendra-1995-ID3040}, increasing differences in the plasma density obtained by the classical LMEA based fluid model and by PIC/MCC simulations
with rising gas pressure (20\,\% at 4\,Pa to 35\,\% at 40\,Pa) have been found in~\cite{Turner-2013-ID3004}.

In Ref.~\cite{Surendra-1993-ID812} moments of the Boltzmann equation have been calculated using PIC/MCC simulations. It has been shown that the convective term in the momentum balance equation for ions is of particular importance if the collisionality of the sheaths is low, while it can be neglected for electrons if secondary electron emission is of minor importance. Furthermore, it has been pointed out that the spatial and temporal variation of the electron energy plays a predominant role in low-pressure ccrf discharges.

Different fluid models for low-pressure ccrf discharges have previously been compared, e.g., by Young \textit{et al.}~\cite{Young-1993-ID2902} and Chen \textit{et al.}~\cite{Chen-2004-ID2132}. It has been shown that the reliability of fluid models for ccrf discharges can be enhanced by an adequate description of electron momentum and energy transport~\cite{Young-1993-ID2902,Chen-2004-ID2132} and, particularly, by the introduction of an integral form for the electron heat flux which provides an improved prediction of electron heating mechanisms~\cite{Chen-2004-ID2132}.

In Refs.~\cite{Becker-2013-ID2934,Becker-2013-ID3118,Becker_dissbook,Becker-2013-ProcICPIGa,Becker-2012-ProcESCAMPIG} it has been pointed out that an improved prediction of electron heating and energy transport mechanisms can also be achieved by the recently developed drift-diffusion model~\cite{Becker-2013-ID2934}.
In the present paper, this fluid modelling approach is described and applied to the analysis of low-pressure ccrf discharges. 
The fluid modelling results are compared to results obtained by PIC/MCC simulations in order to evaluate the applicability range of the improved fluid modelling approach and to demonstrate its advantages against the classical fluid description for low-pressure ccrf discharges. Following the strategy of Turner \textit{et al.}~\cite{Turner-2013-ID3004} and Surendra~\cite{Surendra-1995-ID3040}, well defined benchmark conditions for ccrf discharges in helium and argon are considered for this purpose.

\section{Methods}

\begin{figure}[ht]\centering%
	\includegraphics[clip,width=0.7\linewidth,trim=0 0 2cm 0]{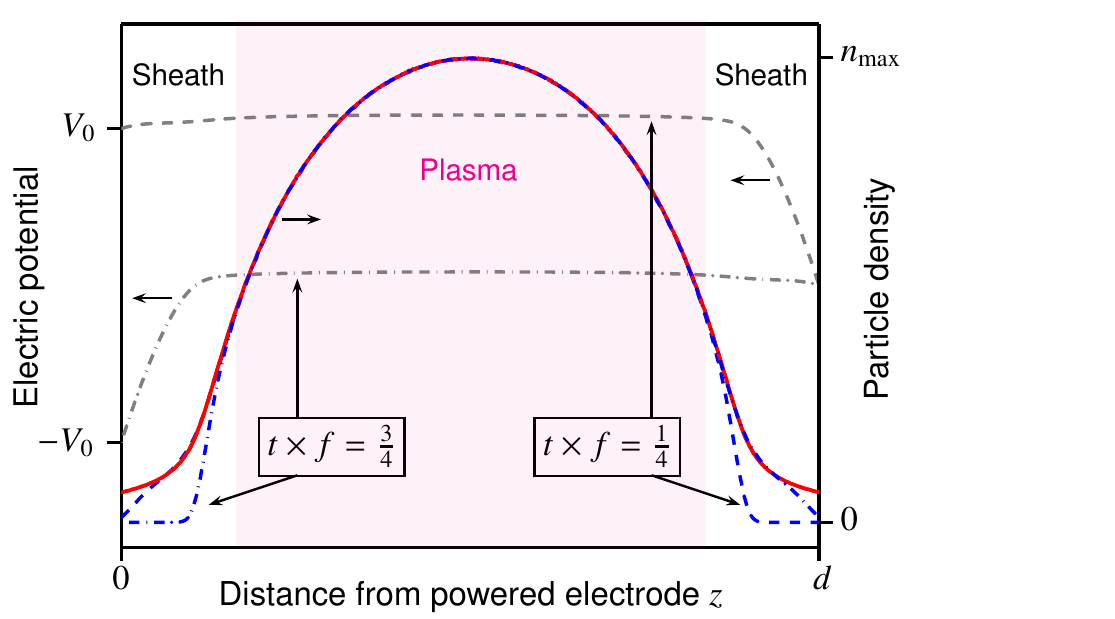}%
	\caption{Sketch of the spatially one-dimensional discharge configuration with time $t$, voltage amplitude $V_0$, frequency $f$, electrode separation $d$ and maximum plasma density $n_\mathrm{max}$. Dashed grey lines: electric potential at two different times $t$; dashed blue lines: electron density at two different times $t$; solid red line: ion density.%
	\label{fig:DischargeConfig}}%
\end{figure}

The evaluation of the applicability range and accuracy of two different fluid modeling approaches for the analysis of low-pressure ccrf discharges is performed considering simple, spatially one-dimensional discharge situations in helium and argon. 
The configuration and main discharge features are sketched in figure~\ref{fig:DischargeConfig}.
The plasma between plane electrodes separated by the distance $d$ is driven by a sinusoidal voltage with amplitude $V_0$ and frequency $f$. 
The specific discharge parameters used for four different cases in helium and three different cases in argon are listed in table~\ref{tab:discharge}.

\begin{table}[ht]\footnotesize\centering
	\caption{Discharge configuration for different cases in helium and argon plasmas. The amplitude of the applied voltage is chosen such that a current density amplitude of about 10\,\nicefrac{A}{m$^2$} is obtained in each case.}
	\label{tab:discharge}
    \begin{tabular}{l l  c c c c  c  c c c}
        \toprule        
		 & & \multicolumn{4}{c}{Helium} & & \multicolumn{3}{c}{Argon}\\ 
		\cmidrule[1.5pt]{3-6} \cmidrule[1.5pt]{8-10}             
		Parameter & Symbol & C1	& C2 & C3 & C4 & & C2& C3 & C4\\		
        \midrule
		Pressure [Pa] & $p$ & 10 & 20 & 40 & 80 & & 20 & 40 & 80\\  		     	
		Gas temperature [K] & $T_\mathrm{gas}$ & \multicolumn{4}{c}{300} & &\multicolumn{3}{c}{300}\\
		Electrode distance [cm] & $d$	& \multicolumn{4}{c}{6.7} & &\multicolumn{3}{c}{2.5}\\		
		Frequency [MHz] & $f$ & \multicolumn{4}{c}{13.56} & &\multicolumn{3}{c}{13.56}\\
		Voltage amplitude [V] & $V_0$ & 250 & 190 & 150 & 125 & & 90 & 70 & 60\\ 			
        \bottomrule
    \end{tabular}
\end{table}

For the conditions considered in the present work, ions cannot follow the oscillating electric field while  the electron density strongly changes with time $t$ in the sheath regions.
The maximum plasma density $n_\mathrm{max}$ typically occurs at the centre of the discharge region. 
As in Ref.~\cite{Turner-2013-ID3004}, it is assumed that the plasma is composed only of electrons and positive ions in the background gas helium or argon. Collision processes are limited to interactions between these charge carriers and the neutral background gas.

The fluid description of the electron component is performed by means of two different drift-diffusion approaches: the novel drift-diffusion model introduced in Ref.~\cite{Becker-2013-ID2934} and abbreviated by DDAn (cf. section~\ref{sec:DDAn}) and the commonly used classical drift-diffusion model, named DDA53, using simplified electron energy transport coefficients~\cite{Rafatov-2012-ID2706} (cf. section~\ref{sec:DDA53}).
Even though elaborations of both approaches have been published elsewhere, specific details of the present implementations are given in sections~\ref{sec:DDAn} and~\ref{sec:DDA53}.
In addition, a time-dependent two-moment model for ions is applied. It takes into account the continuity equation and the momentum balance equation and is described in see section~\ref{sec:ions}. 
The balance equations for electrons and ions are complemented by the Poisson equation
\begin{equation}
	-\frac{\partial^2}{\partial z^2}\phi(z,t) = \frac{e_0}{\varepsilon_0}\Bigl( n_\mathrm{i}(z,t) - n_\mathrm{e}(z,t)\Bigr)
	\label{eq:poisson}
\end{equation}
for the electric potential $\phi$. 
Here, $e_0$, $\varepsilon_0$, $n_\mathrm{i}$ and $n_\mathrm{e}$ denote the elementary charge, the vacuum permittivity and the ion and electron densities, respectively.
Details about the boundary conditions are given in section~\ref{sec:bc} and the numerical solution of the coupled set of fluid model equations is described in section~\ref{sec:numerics_fluid}.

In addition to the fluid modelling approaches, two PIC/MCC simulation codes were developed independently and are applied for mutual verification and benchmarking of the different fluid models for the parameter range considered. Details of the PIC/MCC simulation procedures are given in sections~\ref{sec:pic}.

\subsection{Novel drift-diffusion approximation for electrons}
\label{sec:DDAn}

The particle density $n_\mathrm{e}$ and the energy density $w_\mathrm{e}$  of electrons with mass $m_\mathrm{e}$ are determined by the solution of the coupled balance equations
\begin{eqnarray}
	\frac{\partial}{\partial t}n_\mathrm{e}(z,t)
		+ \frac{\partial}{\partial z} \mathit{\Gamma}_\mathrm{e}(z,t)
		= S(z,t)\,,\label{eq:ne}\\
	\frac{\partial}{\partial t}w_\mathrm{e}(z,t) 
		+ \frac{\partial}{\partial z} Q_\mathrm{e}(z,t)
		= -e_0\mathit{\Gamma}_\mathrm{e}(z,t) E(z,t) - P(z,t)\,.\label{eq:we}
\end{eqnarray}
A consistent drift-diffusion approximation for the particle flux $\mathit{\Gamma}_\mathrm{e}$ and the energy flux $Q_\mathrm{e}$ of the electrons has been deduced by an expansion of the electron velocity distribution function (EVDF) in Legendre polynomials and the derivation of the first four moment equations from the electron Boltzmann equation~\cite{Becker-2013-ID2934,Becker-2013-ID3118}. It reads
\begin{eqnarray}
	\mathit{\Gamma}_\mathrm{e}(z,t)
		= -\frac{1}{m_\mathrm{e}\nu_\mathrm{e}}\frac{\partial}{\partial z}
		\Bigl((\xi_0 + \xi_2)n_\mathrm{e}(z,t) \Bigr)
		-\frac{e_0}{m_\mathrm{e}\nu_\mathrm{e}} E(z,t) n_\mathrm{e}(z,t)\,, \label{eq:JeDDAn}\\
	Q_\mathrm{e}(z,t) = -\frac{1}{m_\mathrm{e}\tilde{\nu}_\mathrm{e}}\frac{\partial}{\partial z}
		\Bigl((\tilde{\xi}_0 + \tilde{\xi}_2) w_\mathrm{e}(z,t) \Bigr) \label{eq:QeDDAn}\\
			\qquad\qquad\; -\frac{e_0}{m_\mathrm{e}\tilde{\nu}_\mathrm{e}}\Bigl(\frac{5}{3}
			+ \frac{2}{3}\frac{\xi_2}{\xi_0}\Bigr) E(z,t) w_\mathrm{e}(z,t) \nonumber
\end{eqnarray}
and includes the momentum and energy flux dissipation frequencies $\nu_\mathrm{e}$ and $\tilde{\nu}_\mathrm{e}$ as well as the transport coefficients $\xi_0$, $\xi_2$, $\tilde{\xi}_0$ and $\tilde{\xi}_2$. These properties are given as integrals of the isotropic part $f_0$ and the first two contributions $f_1$ and $f_2$ to the anisotropy of the EVDF over the kinetic energy $U$ of the electrons, respectively, according to
\begin{eqnarray}
	\nu_\mathrm{e} = \frac{2}{3m_\mathrm{e}\mathit{\Gamma}_\mathrm{e}}\int\limits_0^\infty 
	\frac{U^{\nicefrac{3}{2}}}{\lambda_\mathrm{e}(U)} f_1(U)\,\mathrm{d}U\,, \label{eq:nu}\\
	\tilde{\nu}_\mathrm{e} = \frac{2}{3m_\mathrm{e} Q_\mathrm{e}}\int\limits_0^\infty 
	\frac{U^{\nicefrac{5}{2}}}{\lambda_\mathrm{e}(U)} f_1(U)\,\mathrm{d}U\,, \label{eq:enu}\\
	\xi_0 = \frac{2}{3 n_\mathrm{e}}\int\limits_0^\infty 
	U^{\nicefrac{3}{2}} f_0(U)\,\mathrm{d}U\,, \label{eq:transp_first}\\
	\xi_2 = \frac{4}{15 n_\mathrm{e}}\int\limits_0^\infty 
	U^{\nicefrac{3}{2}} f_2(U)\,\mathrm{d}U\,, \\	
	\tilde{\xi}_0 = \frac{2}{3 n_\mathrm{e}}\int\limits_0^\infty 
	U^{\nicefrac{5}{2}} f_0(U)\,\mathrm{d}U\,, \\
	\tilde{\xi}_2 = \frac{4}{15 n_\mathrm{e}}\int\limits_0^\infty 
	U^{\nicefrac{5}{2}} f_2(U)\,\mathrm{d}U\,. \label{eq:transp_last}		
\end{eqnarray}
Here, $\lambda_\mathrm{e}$ is the electron-energy-dependent mean free path of the electrons~\cite{Grubert-2009-ID2551}.
Their particle density $n_\mathrm{e}$, energy density $w_\mathrm{e}$, particle flux $\mathit{\Gamma}_\mathrm{e}$ and energy flux $Q_\mathrm{e}$ are given by
\begin{eqnarray}
	n_\mathrm{e} = \int\limits_0^\infty U^{\nicefrac{1}{2}} f_0(U) \,\mathrm{d}U\,, \\
	w_\mathrm{e} = \int\limits_0^\infty U^{\nicefrac{3}{2}} f_0(U) \,\mathrm{d}U\,, \\
	\mathit{\Gamma}_\mathrm{e} = \frac{1}{3}\sqrt{\frac{2}{m_\mathrm{e}}} 
		\int\limits_0^\infty U f_1(U)\,\mathrm{d} U\,,\\
	Q_\mathrm{e} = \frac{1}{3}\sqrt{\frac{2}{m_\mathrm{e}}}
	   	\int\limits_0^\infty U^2 f_1(U)\,\mathrm{d} U\,.
\end{eqnarray}
The coefficients~(\ref{eq:nu})--(\ref{eq:transp_last}) are determined within the framework of the common LMEA as functions of the mean electron energy $U_\mathrm{e}=w_\mathrm{e}/n_\mathrm{e}$ by solving the stationary, spatially homogeneous Boltzmann equation in multiterm approximation for different values of the electric field~\cite{Hagelaar-2005-ID2276,Grubert-2009-ID2551}, see section~\ref{sec:inp_phys}.

The source term $S$ in equation~(\ref{eq:ne}) represents the gain of electrons due to ionization of neutral gas atoms in electron-neutral collisions while $P$ in equation~(\ref{eq:we}) describes the loss of electron energy in elastic, exciting and ionizing collisions of electrons with neutral gas atoms. They are given by
\begin{eqnarray}
	S(z,t) = k^\mathrm{io} n_\mathrm{gas} n_\mathrm{e}(z,t)\,,\\
	P(z,t) = \Bigl ( \sum_{j} U^{\mathrm{ex}}_j k^{\mathrm{ex}}_j
		+ U^\mathrm{io} k^\mathrm{io} + \tilde{k}^\mathrm{el}\Bigr) n_\mathrm{gas} n_\mathrm{e}(z,t)\,.
\end{eqnarray}
Here, $n_\mathrm{gas} = p/(k_\mathrm{B}T_\mathrm{gas})$ is the density of the background gas with pressure $p$, temperature $T_\mathrm{gas}$ and mass $M$, $k_\mathrm{B}$ denotes the Boltzmann constant, $k^{\mathrm{ex}}_j$ and $k^\mathrm{io}$ are the respective rate coefficients for excitation and ionization processes with energy thresholds $U^{\mathrm{ex}}_j$ and $U^{\mathrm{io}}$, respectively. 
The energy rate coefficient for energy dissipation in elastic collisions is denoted by $\tilde{k}^\mathrm{el}$. The rate coefficients are given by
\begin{eqnarray}	
	k^{\mathrm{ex}}_j = \frac{1}{n_\mathrm{e}} \sqrt{\frac{2}{m_\mathrm{e}}} \int\limits_0^\infty
		U Q^\mathrm{ex}_j(U) f_0(U)\,\mathrm{d}U\,, \label{eq:k_first}\\
	k^{\mathrm{io}} = \frac{1}{n_\mathrm{e}} \sqrt{\frac{2}{m_\mathrm{e}}} \int\limits_0^\infty
		U Q^\mathrm{io}(U) f_0(U)\,\mathrm{d}U\,, \\	
	\tilde{k}^{\mathrm{el}} = \frac{1}{n_\mathrm{e}} \sqrt{\frac{2}{m_\mathrm{e}}} \int\limits_0^\infty
		2 \, \frac{m_\mathrm{e}}{M} \, U^2 Q^\mathrm{m}(U)\Bigl(f_0(U) 
		+ k_\mathrm{B} T_\mathrm{gas} \frac{\mathrm{d}}{\mathrm{d} U} f_0(U)\Bigr)\,\mathrm{d}U\,,	
	\label{eq:k_last}	
\end{eqnarray}
where $Q^\mathrm{ex}_j$, $Q^\mathrm{io}$ and $Q^\mathrm{m}$ are the electron-neutral collision cross sections for excitation, ionization and momentum transfer in elastic electron-neutral collisions, respectively. 
As for the transport coefficients, the rate coefficients are incorporated into the fluid model as functions of the mean electron energy in the framework of the local mean energy approximation. 

\subsection{Classical drift-diffusion approximation for electrons}
\label{sec:DDA53}

The conventional drift-diffusion model for electrons in nonthermal plasmas comprises the balance equations~(\ref{eq:ne}) and~(\ref{eq:we}) with the expressions
\begin{eqnarray}
	\mathit{\Gamma}_\mathrm{e}(z,t)
		= -\frac{\partial}{\partial z}\Bigl(D_\mathrm{e} n_\mathrm{e}(z,t)\Bigr)
		-b_\mathrm{e} E(z,t) n_\mathrm{e}(z,t)\,, \label{eq:JeDDAc}\\
	Q_\mathrm{e}(z,t) 
		= -\frac{\partial}{\partial z}\Bigl(\tilde{D}_\mathrm{e} w_\mathrm{e}(z,t)\Bigr)
		-\tilde{b}_\mathrm{e} E(z,t) w_\mathrm{e}(z,t) \label{eq:QeDDAc}
\end{eqnarray}
for the particle and energy fluxes~\cite{Hagelaar-2005-ID2276,Grubert-2009-ID2551}. 
The particle and energy diffusion coefficients $D_\mathrm{e}$ and $\tilde{D}_\mathrm{e}$ as well as the mobilities $b_\mathrm{e}$ and $\tilde{b}_\mathrm{e}$ are given by integrals of the EVDF over energy space~\cite{Hagelaar-2005-ID2276,Grubert-2009-ID2551}.
Because numerical problems arise in many situations when the consistent expressions for the energy transport coefficients $\tilde{D}_\mathrm{e}$ and $\tilde{b}_\mathrm{e}$ are used~\cite{Salabas-2003-ID1901,Becker-2013-ID2934}, the simplified expressions $\tilde{D}_\mathrm{e} = 5/3\, D_\mathrm{e}$ and $\tilde{b}_\mathrm{e} = 5/3\, b_\mathrm{e}$ are usually applied~\cite{Hagelaar-2005-ID2276,Rafatov-2012-ID2864,Mihailova-2008-ID2515,Greb-2013-ID3202,Panneer Chelvam-2015-ID3861}. 
This classical drift-diffusion approximation with the transport coefficients
\begin{eqnarray}
	D_\mathrm{e} = \frac{1}{3 n_\mathrm{e}} \sqrt{\frac{2}{m_\mathrm{e}}} \int\limits_0^\infty
		\lambda_\mathrm{e}(U) U  \Bigl(f_0(U) + \frac{2}{5} f_2(U) \Bigr) \,\mathrm{d} U\,, \label{eq:De}\\
	b_\mathrm{e} = -\frac{e_0}{3 n_\mathrm{e}} \int\limits_0^\infty \lambda_\mathrm{e}(U) \Bigl[ 
		U  \frac{\mathrm{d}}{\mathrm{d}U} \Bigl(
		 f_0(U) + \frac{2}{5} f_2(U) \Bigr)  + \frac{3}{5} f_2(U)
		\Bigr]\,\mathrm{d} U \label{eq:be}
\end{eqnarray}
is used here for comparative studies. Note that the simplified energy transport coefficients are valid in case of a Maxwellian EVDF, only.

\subsection{Two-moment model for ions}
\label{sec:ions}

For the description of low-pressure ccrf discharges the ion inertia must be taken into account by considering an effective electric field~\cite{Salabas-2002-ID1762} or by solving the time-dependent momentum balance equation~\cite{Eremin-2016-ID3828}. The latter approach is chosen here and the system of moment equations
\begin{eqnarray}
	\frac{\partial}{\partial t}n_\mathrm{i}(z,t)
		+ \frac{\partial}{\partial z} \mathit{\Gamma}_\mathrm{i}(z,t)
		= S(z,t)\,,\label{eq:ni}\\
	\frac{\partial}{\partial t}\mathit{\Gamma}_\mathrm{i}(z,t)
		+\frac{\partial}{\partial z}\Bigl(\mathit{\Gamma}_\mathrm{i}(z,t) v_\mathrm{i}(z,t)
		+ \frac{p_\mathrm{i}(z,t)}{m_\mathrm{i}} \Bigr) \label{eq:Ji} \\\nonumber
		\qquad\qquad = \frac{e_0}{m_\mathrm{i}} n_\mathrm{i}(z,t) E(z,t) 
		- \nu_\mathrm{i} \mathit{\Gamma}_\mathrm{i}(z,t)
\end{eqnarray}		
is solved to determine the density $n_\mathrm{i}$ and the particle flux $\mathit{\Gamma}_\mathrm{i}$ of ions with mean velocity $v_\mathrm{i} = \mathit{\Gamma}_\mathrm{i} / n_\mathrm{i}$ and mass $m_\mathrm{i}$. The ion pressure $p_\mathrm{i}$ is given by the ideal gas law $p_\mathrm{i} = n_\mathrm{i} k_\mathrm{B}T_\mathrm{i}$ where it is assumed that the heating of ions is negligible, i.e., $T_\mathrm{i}=T_\mathrm{gas}$.
Although the assumption  $T_\mathrm{i}\approx T_\mathrm{gas}$ is generally not valid in the sheath regions of ccrf discharges, it does not have any significant impact since the pressure term in equation~(\ref{eq:Ji}) is generally unimportant~\cite{Surendra-1993-ID812}.
The ion momentum dissipation frequency $\nu_\mathrm{i}$ is obtained from measured ion mobilities $b_\mathrm{i}$ according to $\nu_\mathrm{i}=e_0/(m_\mathrm{i} b_\mathrm{i})$, see section~\ref{sec:inp_phys}.

\subsection{Boundary conditions and initial values}
\label{sec:bc}

The Poisson equation~(\ref{eq:poisson}) is supplemented with the boundary conditions $\phi(0,t)=V_0\,\mathrm{sin}(2\pi f t)$ and $\phi(d,t) = 0$, according to the setup shown in figure~\ref{fig:DischargeConfig}. Note that the desired amplitude $J_0$ of the discharge current density
\begin{equation}
	J(t) = \varepsilon_0 \frac{\mathrm{d}}{\mathrm{d}t}  E(0,t)
		+ e_0 \Bigl(\mathit{\Gamma}_\mathrm{i}(0,t) - \mathit{\Gamma}_\mathrm{e}(0,t) \Bigr)
\end{equation}
is used as input for the fluid modelling and the amplitude $V_0$ of the rf voltage  applied to the powered electrode is automatically adapted in each period according to
\begin{equation}
    V_0^\mathrm{new} = V_0^\mathrm{old}
    \left(\frac{J_0}{J_\mathrm{calc}}\right)
    \label{eq:adaptivevoltage}
\end{equation}
until a periodic state is reached. Here, $J_\mathrm{calc}$ is the actual amplitude of the discharge current density. This procedure ensures that $J_\mathrm{calc} = J_0$ when periodic state is reached. 

In order to exclude any uncertainties regarding the implementation of boundary effects in the different modelling approaches, the boundary conditions for the particles are set as simple as possible. It is assumed that neither reflection of particles nor emission of secondary electrons occur at the electrodes located at $z=0$ and $z=d$. The expression applied at the boundaries for electrons ($j=\mathrm{e}$) and ions ($j=\mathrm{i}$) reads
\begin{equation}
	\mathit{\Gamma}_j\cdot\nu = \Bigl(\mathrm{max}(v^\mathrm{d}_j\cdot\nu ,0) + \frac{1}{4}v^\mathrm{th}_j\Bigr) n_j\,,
	\label{eq:BCparticles}
\end{equation}
where $\nu=-1$ at $z=0$ and $\nu=1$ at $z=d$. The drift velocity $v^\mathrm{d}_j$ is deduced from the respective expressions in equations~(\ref{eq:JeDDAn}),~(\ref{eq:JeDDAc}) and~(\ref{eq:Ji}) and $v^\mathrm{th}_j = \sqrt{8 k_\mathrm{B} T_j/(\pi m_j)}$ denotes the thermal velocity of species $j$. For electrons, the ``temperature'' $T_\mathrm{e} = 2\,U_\mathrm{e}/(3\,k_\mathrm{B})$ is used here.
Similarly, the boundary condition for the electron energy balance equation~(\ref{eq:we}) is given by
\begin{equation}
	Q_\mathrm{e}\cdot\nu = \Bigl(\mathrm{max}(\tilde{v}^\mathrm{d}_\mathrm{e}\cdot\nu ,0) + \frac{1}{3}v^\mathrm{th}_\mathrm{e}\Bigr) w_\mathrm{e}\,,
\end{equation}
where $\tilde{v}^\mathrm{d}_\mathrm{e} = -\frac{e_0}{m_\mathrm{e}\tilde{\nu}_\mathrm{e}}\Bigl(\frac{5}{3} + \frac{2}{3}\frac{\xi_2}{\xi_0}\Bigr) E$ in case of the novel drift-diffusion approximation~(\ref{eq:QeDDAn}) and $\tilde{v}^\mathrm{d}_\mathrm{e} = -5/3\, b_\mathrm{e} E$ for the classical drift-diffusion approximation.

The choice of boundary conditions for the hyperbolic system~(\ref{eq:ni})--(\ref{eq:Ji}) requires special care because the number of required boundary conditions at each boundary depends on the direction of the characteristics as well as the ion sound speed $c_\mathrm{i}=\sqrt{k_\mathrm{B} T_\mathrm{i}/m_\mathrm{i}}$ as described, e.g., in Refs.~\cite{Wesseling2001,Wilcoxson-1996-ID1020}. 
Table~\ref{tab:ionbc} lists the number of boundary conditions that is applied here. The condition $\partial n_\mathrm{i}/\partial z=0$ is used in addition to equation (\ref{eq:BCparticles}) if two boundary conditions are to be set. A logarithmic extrapolation of ion properties is performed to complete the number of physical boundary conditions where required.

\begin{table}[ht]\footnotesize\centering%
	\caption{Number of boundary conditions to be set for the hyperbolic system~(\ref{eq:ni})--(\ref{eq:Ji}) depending on the mean ion velocity $v_\mathrm{i}$ and the ion sound speed $c_\mathrm{i}$.} %
	\label{tab:ionbc} %
    \begin{tabular}{l c c}
        \toprule
		 & $|v_\mathrm{i}| < c_\mathrm{i}$ & $|v_\mathrm{i}| > c_\mathrm{i}$ \\
		\midrule
		$v_\mathrm{i}\cdot\nu < 0$ & 1 & 2 \\
		$v_\mathrm{i}\cdot\nu > 0$ & 1 & 0 \\
        \bottomrule
    \end{tabular}
\end{table}

In addition to boundary conditions, initial values are to be set for the particle, momentum and energy balance equations (\ref{eq:ne}), (\ref{eq:we}), (\ref{eq:ni}) and (\ref{eq:Ji}). Here, a quasi-neutral state with a homogeneous initial density $n_j(z,0) = n_\mathrm{init}$ of charged particles and mean electron energy $U_\mathrm{e}(z,0)=U^\mathrm{init}_\mathrm{e}$ is assumed. The mean ion velocity is set to zero at $t=0$.
Since we do not allow for the emission of secondary electrons from the electrodes, the initial charge-carrier density $n_\mathrm{init}$ must be large enough to sustain the plasma. However, it was verified that the periodic discharge behaviour is not influenced by the specific values used. 

\subsection{Numerical solution of fluid models}
\label{sec:numerics_fluid}

For the numerical solution of the system of nonlinearly coupled equations a finite-differences discretization in space is performed and a semi-implicit time-stepping scheme is applied. The number of spatial grid points and the size of the time step used for the numerical solution are specified below. 
Both parameters are chosen such that converged results are obtained. 

For Poisson's equation~(\ref{eq:poisson}) the standard central difference quotient of second order~\cite{thomas_book} is used while the parabolic drift-diffusion equations (\ref{eq:ne}) and (\ref{eq:we}) for the particle and energy density of electrons with fluxes (\ref{eq:JeDDAn}) and (\ref{eq:QeDDAn}) or (\ref{eq:JeDDAc}) and (\ref{eq:QeDDAc}), respectively, are discretized by means of the exponentially fitted Scharfetter-Gummel finite-difference scheme~\cite{Scharfetter-1969-ID153} as described in Ref.~\cite{Becker-2009-ID2678}. 
The hyperbolic system of moment equations (\ref{eq:ni}) and (\ref{eq:Ji}) for the density  and the flux of ions is discretized in space by the standard first-order upwind scheme~\cite{thomas_book} and a predictor-corrector approach is chosen for the time-coupling of the equations as detailed in Ref.~\cite{Becker_dissbook}.
Here, the number of boundary conditions affects the solution method by evaluating the ion sound speed and the mean ion velocity for each time step and setting the required physical and numerical boundary conditions according to table~\ref{tab:ionbc}.
The correctness of the computer code has previously been verified by the comparison with other methods and computer codes for a wide range of test problems and discharge conditions~\cite{Becker_dissbook,Derzsi-2009-ID2552,Alili-2016-ID4037}.

\subsection{PIC/MCC simulation procedure}
\label{sec:pic}

Two independent PIC/MCC simulation codes for low-pressure ccrf plasmas were developed at the Institute of Theoretical Physics and Astrophysics (ITAP), University of Kiel, Germany (named PIC(ITAP)) and at the Leibniz Institute for Plasma Science and Technology (INP) Greifswald, Germany (named PIC(INP)) for mutual verification and for validation of the fluid models.
Details of the PIC(INP) method, which has been extended from a previous model for streamer simulations~\cite{Teunissen-2014-ID3360},
are given in Ref.~\cite{Sun-2016-ID3996}.
Although the general procedure of both PIC/MCC codes is the same a brief description is given here.

Compared to the fluid models discussed above, PIC/MCC simulations 
are particle-based, i.e., they track the trajectories of so-called 
superparticles under the influence of the self-consistent electric field 
determined by solving equation~(\ref{eq:poisson}). 
The two PIC/MCC simulation codes used 
here resolve one space dimension and trace all three velocity 
components, usually referred to as 1d3v. Particles are represented in 
the cloud-in-cell scheme, where the weight factor for the charge 
distribution on the grid decreases linearly from the particle position 
toward the grid points. 
The weight of superparticles  is constant in all simulations, i.e., the adaptive particle management available in the PIC(INP) code is disabled to avoid possible uncertainties.
Both PIC/MCC codes are parallelized using the Message Passing Interface (MPI).

For the time integration of the equations of 
motion the velocity Verlet algorithm is used as opposed to the leapfrog 
method usually applied in many other codes~\cite{Turner-2013-ID3004}. 
It has been shown in Ref.~\cite{Sun-2016-ID3996} that the velocity Verlet method converges faster than the leapfrog scheme with respect to the size of the time step used in PIC/MCC simulations of low-pressure ccrf discharges.

Particles that reach the electrode surface at $z=0$ or $z=d$ are fully absorbed and are 
removed from the simulation. 
This is consistent with the boundary conditions applied 
for the fluid models (see section~\ref{sec:bc}).
Except for backscattering in elastic ion-neutral collisions, the particles are scattered isotropically 
in the collision events and the remaining energy in ionizing
collisions is shared 
in equal parts between the two electrons. Instead of 
calculating the collision probability for all particles individually, 
the null-collision method is used. Further details on the PIC/MCC algorithm can be found, e.g., in Refs.~\cite{Donko-2011-ID2668,birdsall_book,Sun-2016-ID3996}.
Table~\ref{tab:methods} summarizes the applied fluid and PIC/MCC simulation methods used for the comparative investigations.

\begin{table}[ht]\footnotesize\centering
	\caption{Applied fluid and PIC/MCC simulation methods.}
	\label{tab:methods}
    \begin{tabular}{l p{11cm}}
        \toprule         
		Method & Description \\		
        \midrule
		DDAn & Continuity equations (\ref{eq:ne}), (\ref{eq:we}) for electrons using drift-diffusion fluxes (\ref{eq:JeDDAn}), (\ref{eq:QeDDAn}); particle and momentum balance equations (\ref{eq:ni}), (\ref{eq:Ji}) for ions;\\
		DDA53 & Continuity equations (\ref{eq:ne}), (\ref{eq:we}) for electrons using drift-diffusion fluxes (\ref{eq:JeDDAc}), (\ref{eq:QeDDAc}) with $\tilde{D}_\mathrm{e} = 5/3\, D_\mathrm{e}$ and $\tilde{b}_\mathrm{e} = 5/3\, b_\mathrm{e}$; particle and momentum balance equations (\ref{eq:ni}), (\ref{eq:Ji}) for ions;\\
		PIC(ITAP) & PIC/MCC simulation code developed at ITAP Kiel, Germany;\\
		PIC(INP) & PIC/MCC simulation code~\cite{Sun-2016-ID3996} of INP Greifswald, Germany;\\
        \bottomrule
    \end{tabular}
\end{table}

\section{Input data}

\subsection{Physical data}
\label{sec:inp_phys}

Since the aim of this work is to verify the reliability of two different fluid approaches in comparison with PIC/MCC simulations, the sources of uncertainties are reduced by making the physical discharge model of the considered low-pressure ccrf discharges in helium and argon as simple as possible.
Table~\ref{tab:reactions} lists the considered collision processes with their energy thresholds where applicable and the sources from which the corresponding cross section data sets are taken.  

\begin{table}[ht]\footnotesize\centering
	\caption{Collision processes considered for modelling of low-pressure ccrf discharges in helium and argon with references to the source of cross section data. The ion-atom collisions are relevant for the PIC/MCC simulations, only.}
	\label{tab:reactions}
    \begin{tabular}{l l r l}
        \toprule
		Reaction & Type & Energy  & Reference\\
		& & threshold [eV] \\
		\midrule
		\textit{Helium} \\
		$\mathrm{He + e \rightarrow He + e}$ & Elastic collision & - & \cite{Turner-2013-ID3004, Biagi7p1}\\
		$\mathrm{He + e \rightarrow He^* + e}$ & Excitation (triplet) & 19.82 & \cite{Turner-2013-ID3004, Biagi7p1}\\
		$\mathrm{He + e \rightarrow He^{**} + e}$ & Excitation (singlet) & 20.61 & \cite{Turner-2013-ID3004, Biagi7p1}\\
		$\mathrm{He + e \rightarrow He^+ + 2e}$ & Ionization & 24.59 & \cite{Turner-2013-ID3004, Biagi7p1}\\
		$\mathrm{He + He^+ \rightarrow He^+ + He}$ & Elastic (backward) & - & \cite{Turner-2013-ID3004, PhelpsHePlusCross}\\
		$\mathrm{He + He^+ \rightarrow He + He^+}$ & Elastic (isotropic) & - & \cite{Turner-2013-ID3004, PhelpsHePlusCross}\\[.2cm]
		\textit{Argon} \\
		$\mathrm{Ar + e \rightarrow Ar + e}$ & Elastic collision & - & \cite{Phelps-1999-ID2920}\\
		$\mathrm{Ar + e \rightarrow Ar^* + e}$ & Excitation & 11.55 & \cite{Phelps-1999-ID2920}\\
		$\mathrm{Ar + e \rightarrow Ar^+ + 2e}$ & Ionization & 15.76 & \cite{Phelps-1999-ID2920}\\
		$\mathrm{Ar + Ar^+ \rightarrow Ar^+ + Ar}$ & Elastic (backward) & - & \cite{Phelps-1994-ID898}\\
		$\mathrm{Ar + Ar^+ \rightarrow Ar + Ar^+}$ & Elastic (isotropic) & - & \cite{Phelps-1994-ID898}\\
        \bottomrule
    \end{tabular}
\end{table}

\begin{figure}[ht]\centering%
	\includegraphics[clip,width=0.47\linewidth]{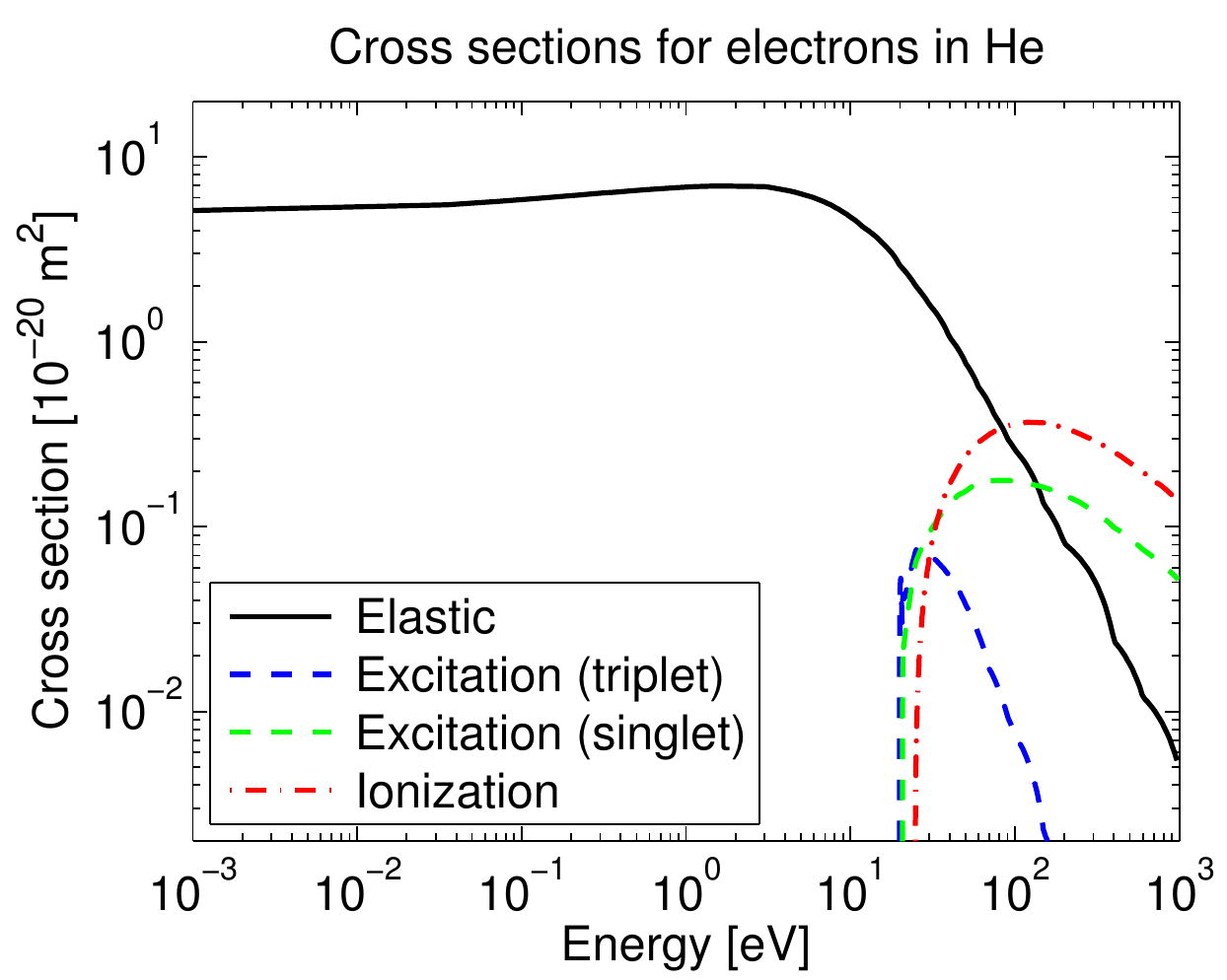}~{\small a)}%
	\includegraphics[clip,width=0.47\linewidth]{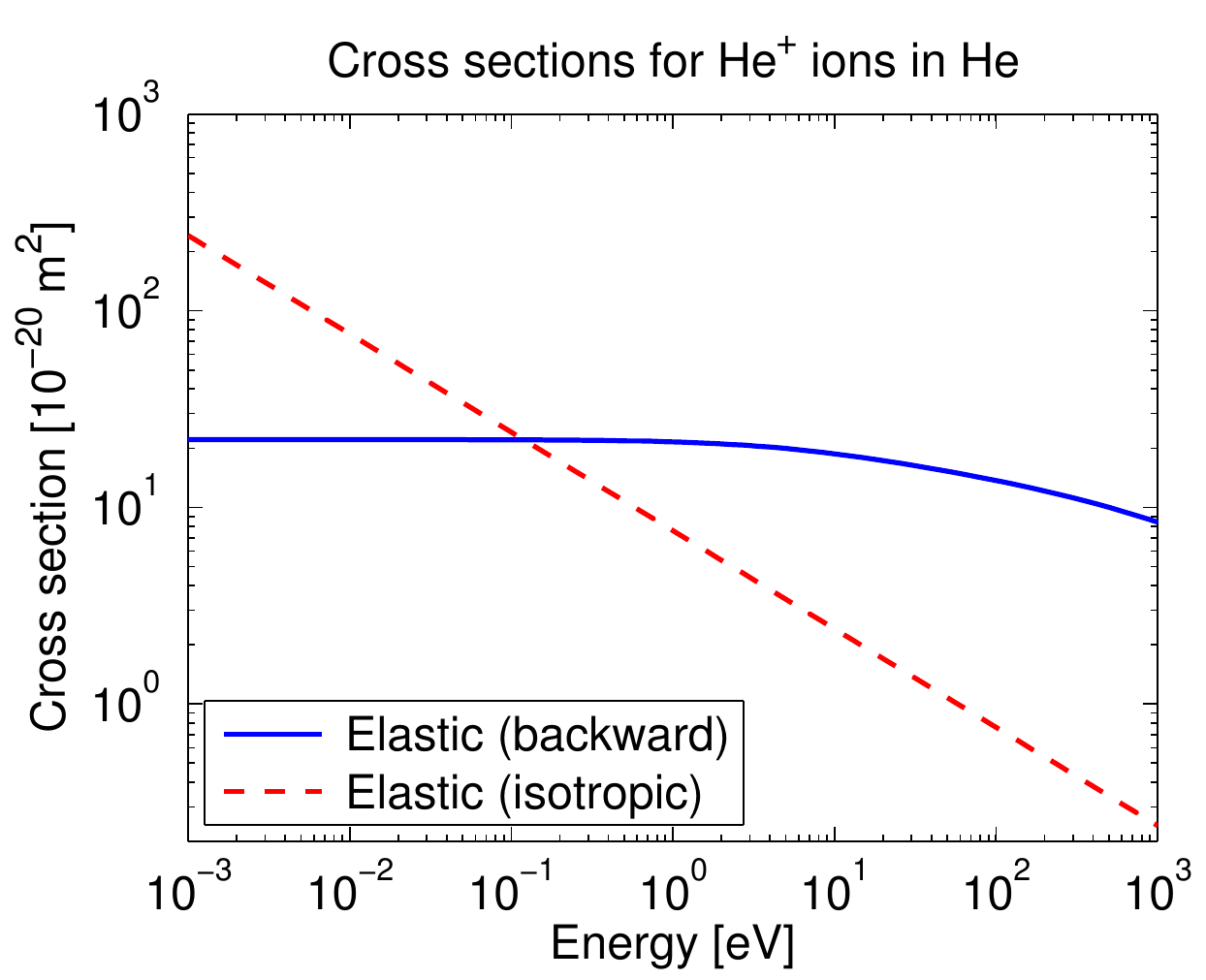}~{\small b)}%
	\caption{Electron-neutral (a) and ion-neutral (b) collision cross sections for helium. The electron energy is in the laboratory frame while for ions the centre-of-mass energy is used as reference.%
	\label{fig:Cross_He}}%
\end{figure}

For helium, the same set of collision processes and cross sections as in Ref.~\cite{Turner-2013-ID3004} is used which include elastic electron-neutral collisions, excitation of the triplet and singlet helium states, direct ionization of helium in its ground state and elastic ion-neutral collisions. The input data used for the present calculations are depicted in figure~\ref{fig:Cross_He}.
Note that the ion cross-sections are used for the PIC/MCC simulations only and are not considered for the fluid modelling.
There, the collisional impact of ion-neutral collisions is taken into account by the momentum dissipation frequency $\nu_\mathrm{i}$ as obtained from measured ion mobilities (cf. section~\ref{sec:ions}).

\begin{figure}[ht]\centering%
	\includegraphics[clip,width=0.47\linewidth]{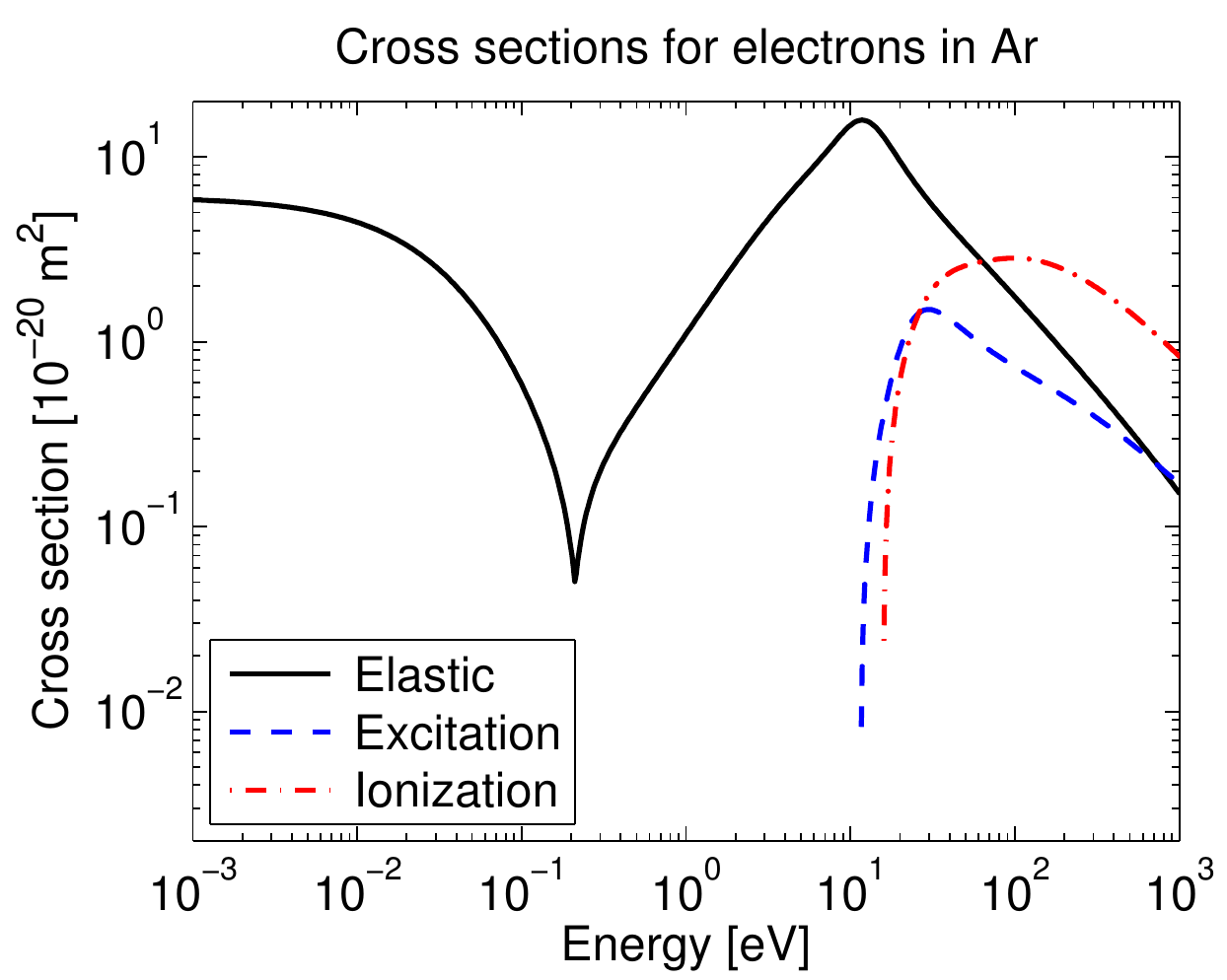}~{\small a)}%
	\includegraphics[clip,width=0.47\linewidth]{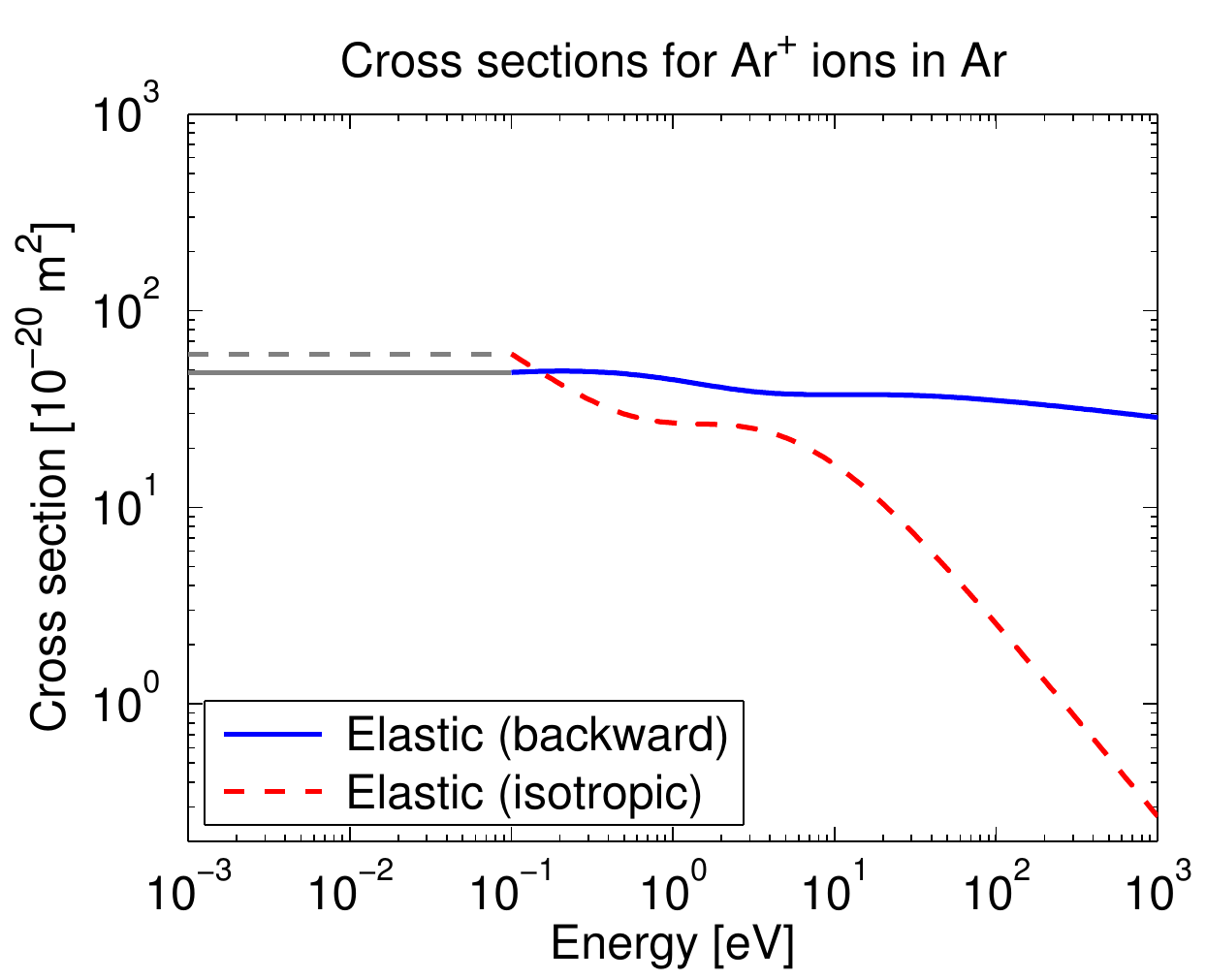}~{\small b)}\\%
	\caption{Electron (a) and ion (b) cross sections for argon as functions of the particle energy in the laboratory frame.%
	\label{fig:Cross_Ar}}%
\end{figure}

A similar set of processes is considered for argon. It comprises elastic electron-neutral and ion-neutral collisions as well as the total electron impact excitation with an energy loss of 11.55\,eV and ionization of argon. As, e.g., in Refs.~\cite{Wilczek-2015-ID3987,Schulze-2015-ID4010} the cross section data set from the JILA database of Phelps~\cite{PhelpsHePlusCross} is used here. More specifically, the fit formulas given in Refs.~\cite{ Phelps-1999-ID2920,Phelps-1994-ID898} with a lower limit of 0.1\,eV for the ion energy are applied for the generation of the cross section input data shown in figure~\ref{fig:Cross_Ar}.

The electron-neutral collision cross sections represented in figures~\ref{fig:Cross_He}a and~\ref{fig:Cross_Ar}a are not only used as input for the PIC/MCC simulations but are additionally utilized for calculation of the momentum and energy flux dissipation frequencies~(\ref{eq:nu}) and~(\ref{eq:enu}), the transport coefficients (\ref{eq:transp_first})--(\ref{eq:transp_last}), (\ref{eq:De}) and (\ref{eq:be}) as well as the rate coefficients (\ref{eq:k_first})--(\ref{eq:k_last}). This is done by solving the stationary, spatially homogeneous Boltzmann equation in eight-term approximation for the required range of electric field strengths. 
Here, the method described in Ref.~\cite{Leyh-1998-ID1222}, which has been modified to take into account nonconservative electron collision processes correctly, has been employed.
For utilization in the fluid models, the obtained transport and rate  coefficients are tabulated as functions of the mean electron energy  which is also obtained by the  solution of the electron Boltzmann equation. 

The ion momentum dissipation frequency appearing in equation~(\ref{eq:Ji}) is determined from measured ion mobilities $b_\mathrm{i}$ depending on the reduced electric field $E/n_\mathrm{gas}$ according to~$\nu_\mathrm{i}=e_0/(m_\mathrm{i} b_\mathrm{i})$. For helium, the $\mathrm{He^+}$ ion mobility is taken from the measurements of Patterson~\cite{Patterson-1970-ID3008} and the measured data of Ellis \textit{et al.}~\cite{Ellis-1976-ID3079} is used for $\mathrm{Ar^+}$.

\subsection{Numerical parameters and initial values}
\label{sec:inp_numerics}

In order to ensure the general validity of the conclusion to be drawn from the present investigations, the considered pressure range for the ccrf discharges in helium and argon is extended as much as possible. For helium, four different pressures (cases C1--C4) in the range from 10 to 80\,Pa are considered (see table~\ref{tab:discharge}). At lower pressures the validity of the drift-diffusion approximation used for the fluid description of electrons becomes inadequate. On the other hand, at pressures above 80\,Pa the PIC/MCC simulations require long computing times and hence they become less appropriate. 

The pressure range used for argon spreads from 20 to 80\,Pa (see table~\ref{tab:discharge}). Here, no converged PIC/MCC simulation results could be obtained for 10\,Pa because of a strong sensitivity of the simulation results on the initial superparticle number $N_\mathrm{sp}$. As pointed out by Turner~\cite{Turner-2006-ID2372,Turner-2013-ID3993}, this effect is caused by velocity space diffusion and cannot easily be circumvented in certain situations.

The numerical input parameters and initial values used for the fluid modelling and PIC/MCC simulations for the four discharge cases in helium and three cases in argon are given in table~\ref{tab:numericalparameters}. 
\begin{table}[ht]\footnotesize\centering
	\caption{Numerical input parameters and initial value used for the simulation of ccrf discharges in helium and argon at a pressure of 10 (C1), 20 (C2), 40 (C3) and 80\,Pa (C4) and a gas temperature of 300\,K.}
	\label{tab:numericalparameters}
    \begin{tabular}{l l  c c c c  c  c c c c}
        \toprule
		 & & \multicolumn{4}{c}{Fluid} &		& \multicolumn{4}{c}{PIC/MCC}\\ 
		\cmidrule[1.5pt]{3-6} \cmidrule[1.5pt]{8-11}         
		Parameter & Symbol & C1& C2 & C3 & C4 &		& C1& C2 & C3 & C4 \\		
        \midrule
		\textit{Helium} \\
        Time steps per period [$10^3$] & $N_\mathrm{\Delta t}$
        	& \multicolumn{4}{c}{4} & 	& \multicolumn{4}{c}{10} \\		
        Number of grid points & $N_\mathrm{\Delta x}$
        	& \multicolumn{4}{c}{671} & 	& \multicolumn{4}{c}{500}   \\ 
        Number of superparticles [$10^5$] & $N_\mathrm{sp}$
        	& \multicolumn{4}{c}{-} & 		& 7.5 & 7.5 & 5.0 & 2.5  \\        	
        Plasma density [$\mathrm{10^{14}\,m^{-3}}$] & $n_\mathrm{init}$
        	& 1.0 & 4.0 & 6.0 & 6.0 & 		& \multicolumn{4}{c}{4.0} \\
        Mean electron energy [eV]  & $U^\mathrm{init}_\mathrm{e}$
        	& \multicolumn{4}{c}{3.88} & 	& \multicolumn{4}{c}{3.88} \\ 
        Mean ion energy [eV]  &  $U^\mathrm{init}_\mathrm{i}$
        	& \multicolumn{4}{c}{-} & 	& \multicolumn{4}{c}{0.039}           	       	
		\\[.2cm]
		\textit{Argon} \\
        Time steps per period [$10^3$] & $N_\mathrm{\Delta t}$
        	& \multicolumn{4}{c}{4} & 	& - & 10 & 10 & 20 \\		
        Number of grid points & $N_\mathrm{\Delta x}$
        	& \multicolumn{4}{c}{501} & 	& - & \multicolumn{3}{c}{500} \\ 
        Number of superparticles [$10^5$] & $N_\mathrm{sp}$
        	& \multicolumn{4}{c}{-} & 		& - & 7.5 & 7.5 & 15 \\           	
        Plasma density [$\mathrm{10^{14}\,m^{-3}}$] & $n_\mathrm{init}$
        	& 1.0 & 4.0 & 6.0 & 6.0 & 		& - & \multicolumn{3}{c}{40.0} \\	
        Mean electron energy [eV]  & $U^\mathrm{init}_\mathrm{e}$
        	& \multicolumn{4}{c}{3.88} & 	& - & \multicolumn{3}{c}{3.88} \\
        Mean ion energy [eV]  &  $U^\mathrm{init}_\mathrm{i}$
        	& \multicolumn{4}{c}{-} & 		& - &\multicolumn{3}{c}{0.039} \\            	
        \bottomrule
    \end{tabular}
\end{table}
Note that special care was placed on choosing the numerical parameters adequately to obtain converged results.
Nevertheless, as discussed in Ref.~\cite{Turner-2013-ID3004}, the accuracy of PIC/MCC simulation results remains difficult to assess as it depends on several numerical parameters and their combination. The development of a generally accepted effective procedure for successively refining the results would be highly desirable.
It is also worth mentioning that the initial plasma density $n_\mathrm{init}$ and energies $U_\mathrm{e,i}^\mathrm{init}$ 
might affect the number of rf periods required to obtain periodic results but
do not influence the periodic behaviour investigated in the following section.
About 1000 (C1) to 4000 (C4) rf periods were required to reach the periodic state for the conditions given in table~\ref{tab:numericalparameters}.

\section{Results and discussion}

Fluid and PIC/MCC calculations are performed for a gas pressure of 10, 20, 40 and 80\,Pa in helium and 20, 40 and 80\,Pa in argon. The amplitude of the applied voltage is chosen such that the same discharge current amplitude of about 10\,\nicefrac{A}{m$^2$} is obtained for each pressure. The respective values and other discharge parameters are given in table~\ref{tab:discharge}.
Prior to the evaluation of the two different fluid modelling approaches DDAn and DDA53 in section~\ref{sec:resFluidPIC}, the discharge behaviour predicted by the PIC/MCC simulation procedures is discussed in the following section.

\subsection{Main discharge features}
\label{sec:resFeatures}

\begin{figure}[ht]\centering%
	\includegraphics[clip,width=0.45\linewidth]{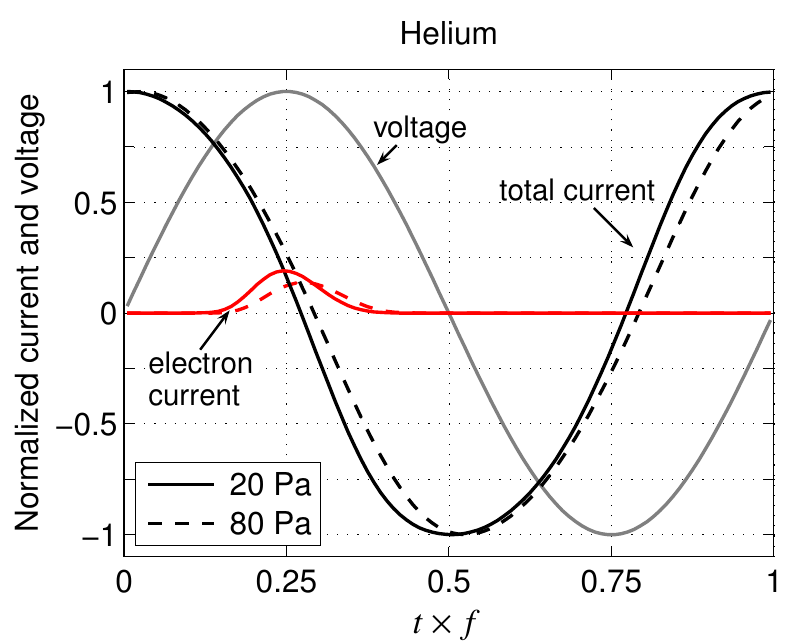}{\small~a)}\hfill%
	\includegraphics[clip,width=0.45\linewidth]{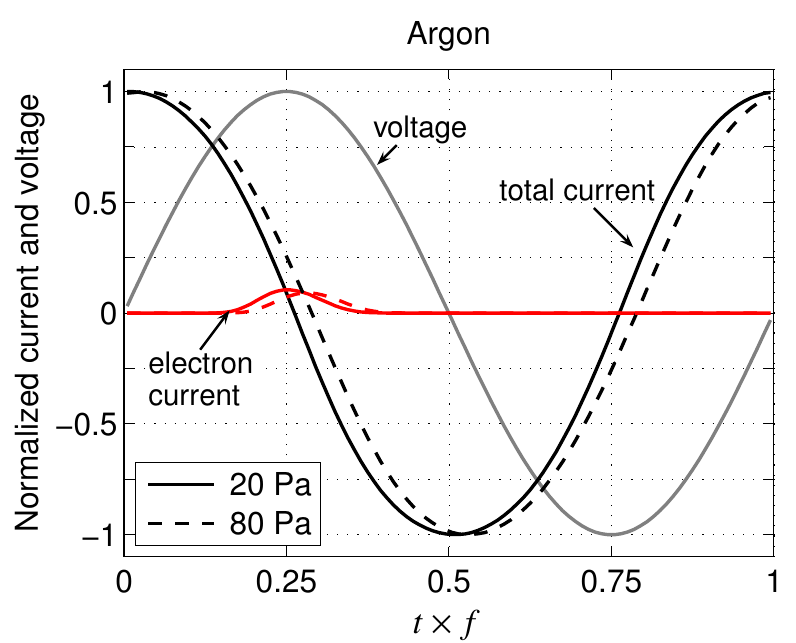}{\small~b)}\\%
	\includegraphics[clip,width=0.47\linewidth]{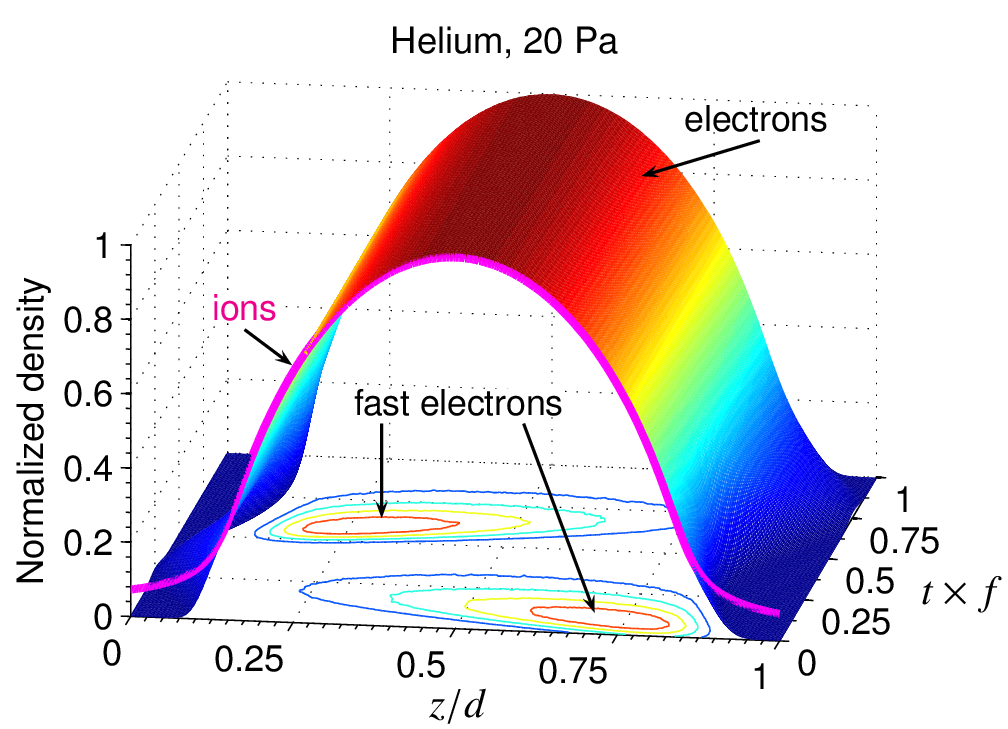}{\small~c)}\hfill%
	\includegraphics[clip,width=0.47\linewidth]{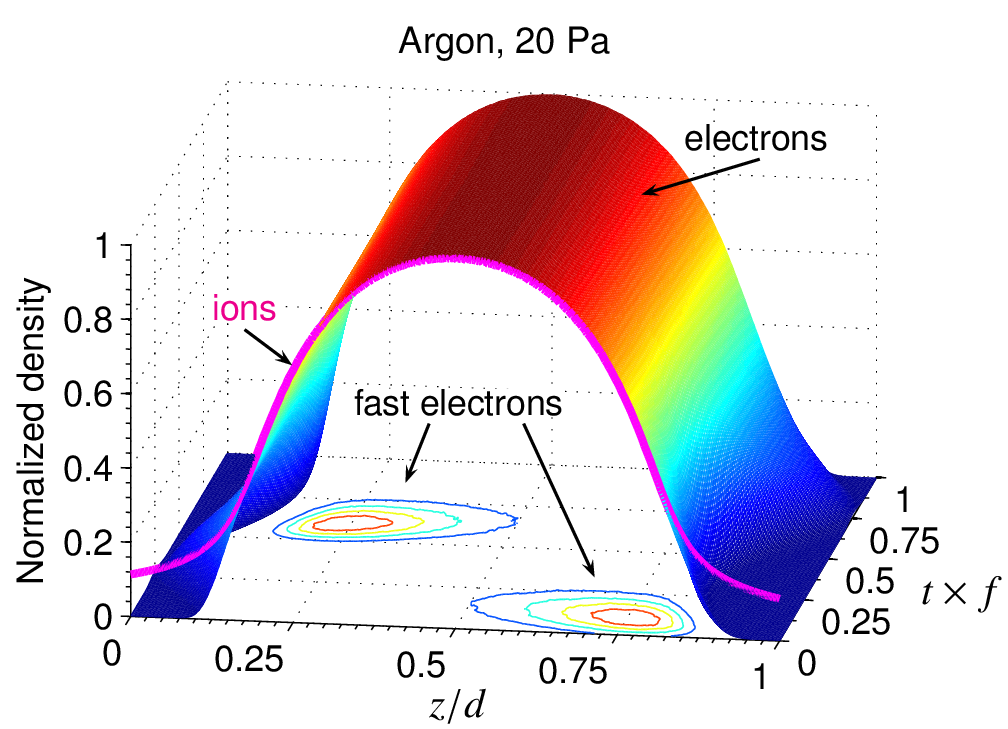}{\small~d)}\\%
	\includegraphics[clip,width=0.47\linewidth]{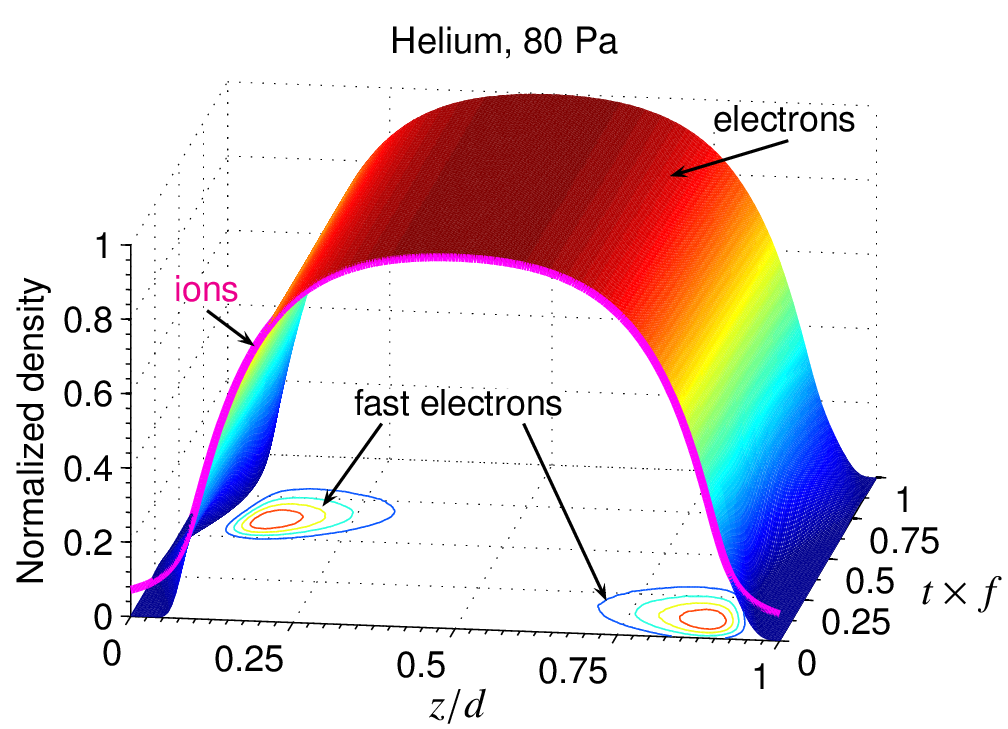}{\small~e)}\hfill%
	\includegraphics[clip,width=0.47\linewidth]{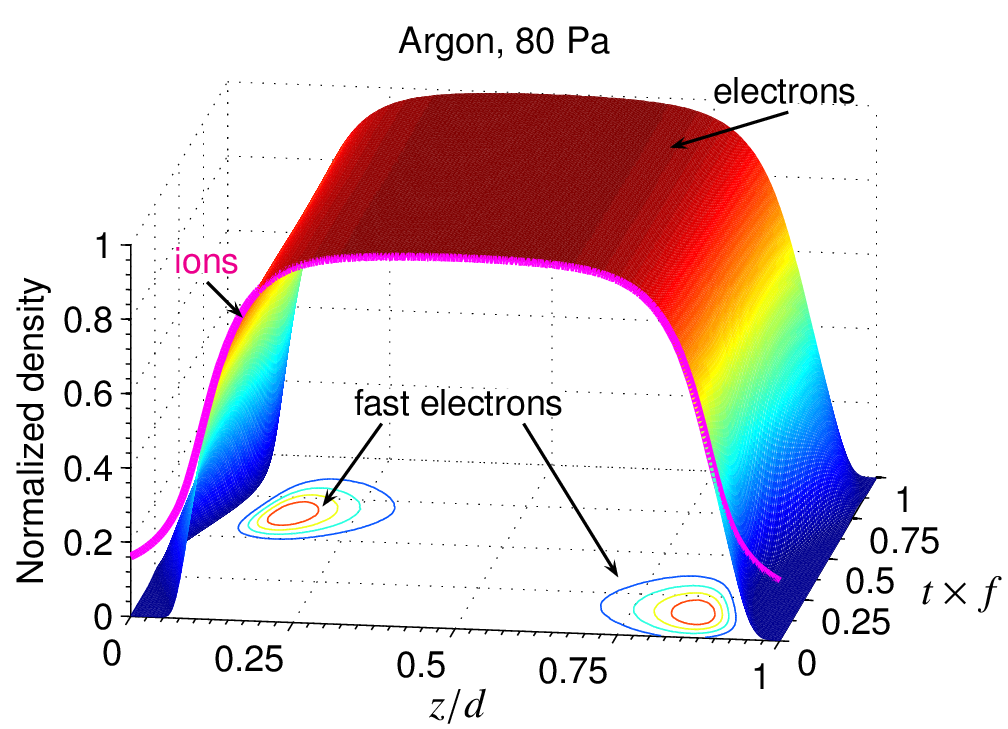}{\small~f)}%
	\caption{Normalized applied  voltage and current density (a,b) and normalized particle densities (c--f) at 20 and 80\,Pa in helium (a,c,e) and argon (b,d,f). 
	\label{fig:plasmaparameters}}%
\end{figure}

The temporal change of the applied voltage and the discharge current density at the powered electrode as well as the spatiotemporal behaviour of the electron and ion densities at 20 and 80\,Pa in helium and argon are depicted in figure~\ref{fig:plasmaparameters}. Here, normalized data are shown. The normalization factors are given in table~\ref{tab:plasmaparameters}.
\begin{table}\footnotesize\centering
	\caption{Discharge parameters obtained by PIC/MCC simulations for ccrf discharges in helium and argon at a gas pressure of 10 (C1), 20 (C2), 40 (C3) and 80\,Pa (C4).}
	\label{tab:plasmaparameters}
    \begin{tabular}{l l r r r r}
        \toprule
        & & \multicolumn{4}{c}{Value} \\
        \cmidrule[1.5pt]{3-6}
		Parameter & Symbol & C1 & C2 & C3 & C4 \\
		\midrule		
		\textit{Helium} \\
		Current density [\nicefrac{A}{m$^2$}] & $J_0$ & 10.0 & 10.0 & 10.1 & 10.6\\
		Plasma density [$\mathrm{10^{15}\, m^{-3}}$] & $n_\mathrm{max}$ & 0.75 & 1.29 & 1.86 & 2.26 \\
		Fast electron density [$\mathrm{10^{12}\, m^{-3}}$] & $n^\mathrm{fast}_\mathrm{e}$ & 0.14 & 0.29 & 1.04 & 3.15\\
		Average ion flux [$\mathrm{10^{18}\, m^{-2} s^{-1}}$] & $\mathit{\Gamma}^\mathrm{avg}_\mathrm{i}$ & 1.59 & 1.44 & 1.28 & 1.20
		\\[.2cm]
		\textit{Argon} \\
		Current density [\nicefrac{A}{m$^2$}] & $J_0$ & - & 10.1 & 10.1 & 10.9\\
		Plasma density [$\mathrm{10^{15}\, m^{-3}}$] & $n_\mathrm{max}$ & - & 1.90 & 2.16 & 2.59\\
		Fast electron density [$\mathrm{10^{12}\, m^{-3}}$] & $n^\mathrm{fast}_\mathrm{e}$ & - & 0.67 & 4.03 & 14.5\\
		Average ion flux [$\mathrm{10^{18}\, m^{-2} s^{-1}}$] & $\mathit{\Gamma}^\mathrm{avg}_\mathrm{i}$ & - & 0.74 & 0.69 & 0.71\\		
        \bottomrule
    \end{tabular}
\end{table}

For all considered conditions ions cannot follow the electric field and hence their density is stationary in time. In contrast, electrons respond almost instantaneously to the change of the electric field in the sheath regions. Hence, the electron current at the momentary anode is maximal close the instant of the largest applied voltage as shown in figures~\ref{fig:plasmaparameters}a,\,b. These figures also show that the phase shift between current and voltage decreases with increasing pressure both in helium and in argon. 
Because the  collision frequency increases with raising pressure, the width of the sheath regions is smaller and the plasma density is larger at 80\,Pa than at 20\,Pa 
(figures~\ref{fig:plasmaparameters}c--f and table~\ref{tab:plasmaparameters}).

\begin{figure}[ht]\centering%
	\includegraphics[clip,width=0.47\linewidth]{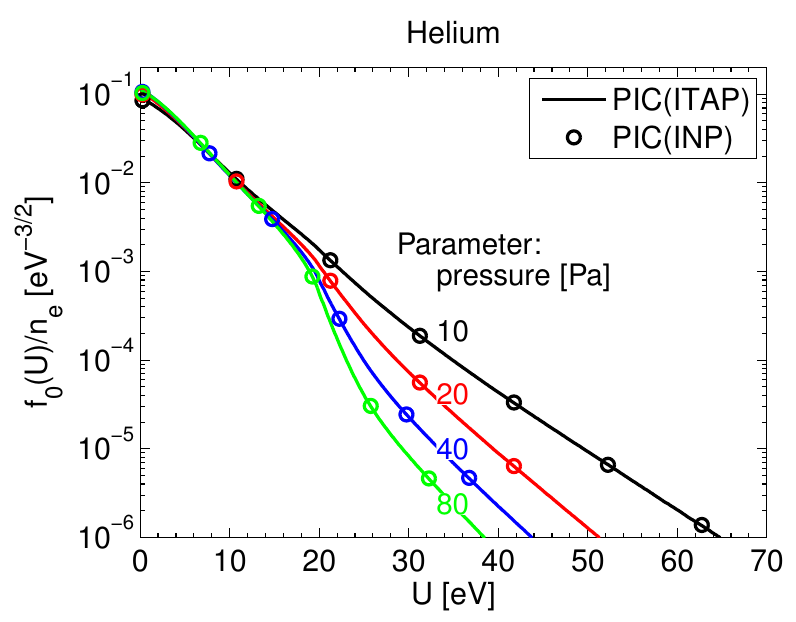}{\small~a)}\hfill%
	\includegraphics[clip,width=0.47\linewidth]{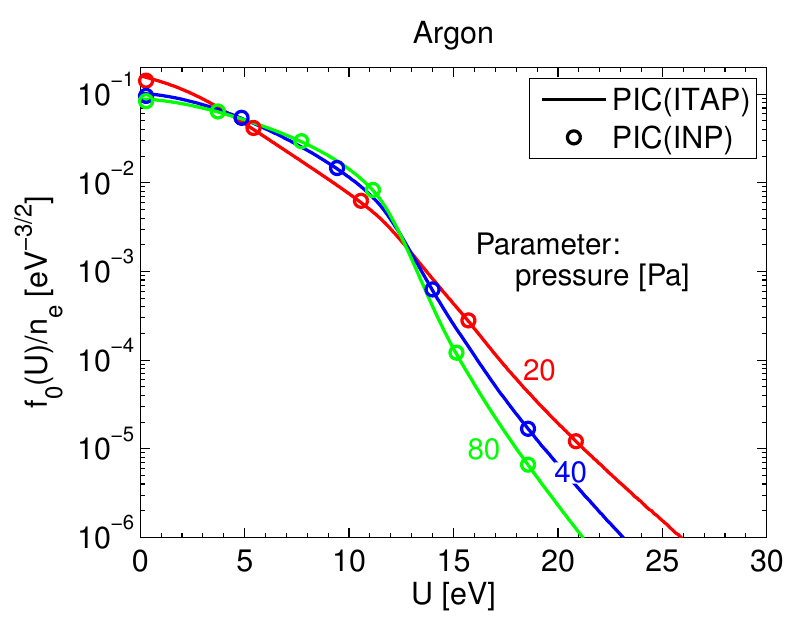}{\small~b)}%
	\caption{Time and space averaged isotropic distribution $f_0/n_\mathrm{e}$ at different pressure in helium (a) and argon (b) obtained by the present PIC/MCC simulation codes. 
	\label{fig:EEDF}}%
\end{figure}

The comparison of the spatiotemporal change of the density of highly energetic electrons in helium (figures~\ref{fig:plasmaparameters}c,\,e) and in argon (figures~\ref{fig:plasmaparameters}d,\,f) reveals certain differences. 
Here,  electrons are considered to be ``fast'' if their energy is larger than $e_0 V_0/4$, where $V_0$ is the voltage amplitude given in table~\ref{tab:discharge}.
In both gases the maximum density of fast electrons occurs in front of the momentary cathode just before the applied voltage reaches its maximum at $t\times f=0.25$ (positive applied voltage, cathode at $z=d$) and at $t\times f=0.75$ (negative applied voltage, cathode at $z=0$), respectively. 
However, in argon the profile of the fast electron density is much more localized than in helium which indicates that in argon the electrons loose their energy more rapidly when propagating towards the anode. 
%
The corresponding time and space averaged isotropic part of the EVDF $f_0(U)/n_\mathrm{e}$ normalized by the electron density $n_\mathrm{e}$ is shown in figure~\ref{fig:EEDF}.
Obviously, it is increasingly influenced by inelastic electron-neutral collisions for larger pressures. The electron impact excitation and ionization processes lead to a marked depletion of the electron population above the lowest threshold energy for exciting collisions, i.e. 19.82\,eV for helium and 11.55\,eV for argon.
The agreement of the two independently developed different PIC/MCC simulation codes for all considered discharge conditions illustrated in figure~\ref{fig:EEDF} and also found for all macroscopic properties mutually verifies their correctness.

\subsection{Comparison of fluid modelling and PIC/MCC simulation results}
\label{sec:resFluidPIC}

\begin{figure}[ht]\centering%
	\includegraphics[clip,width=0.47\linewidth]{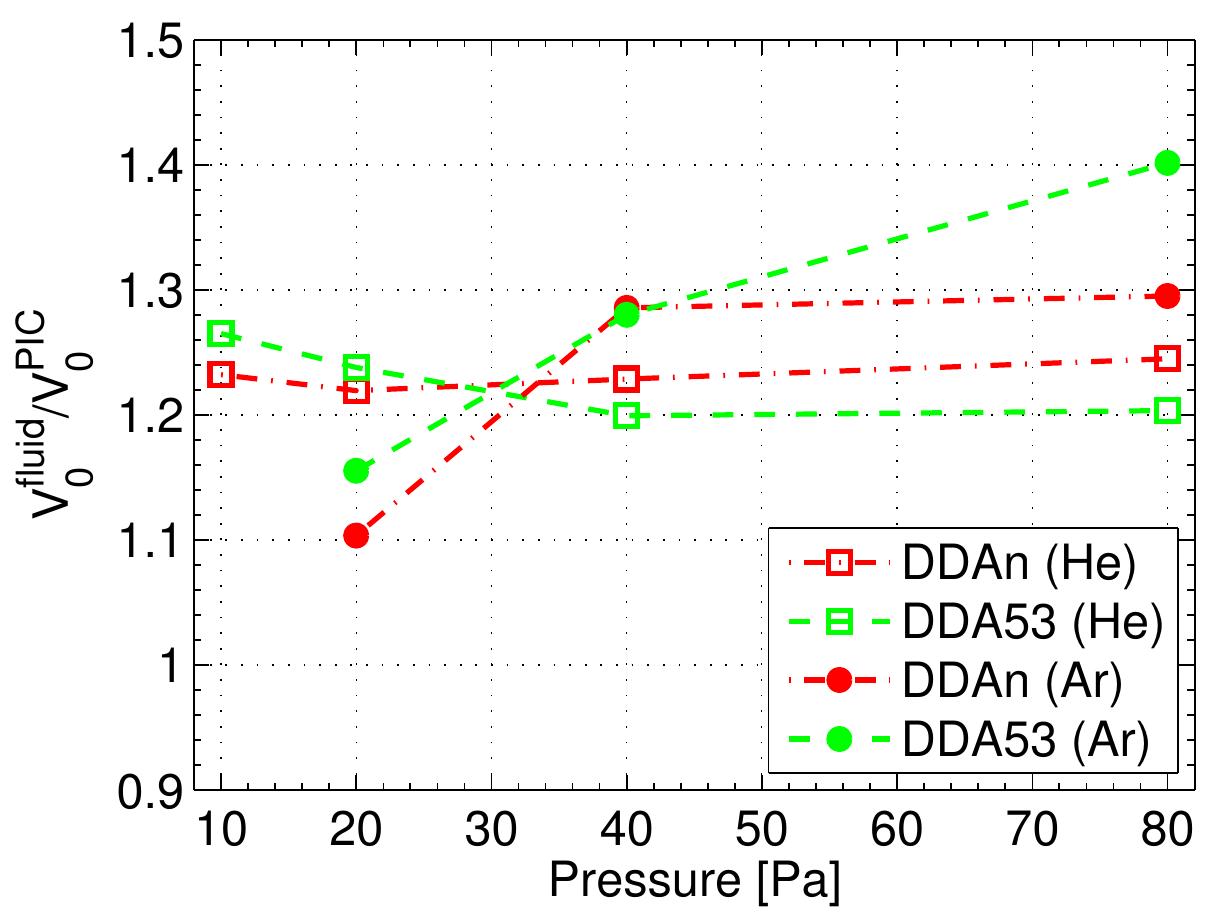}
	\caption{Amplitude of the rf voltage obtained by DDAn and DDA53 in helium and argon as a function of gas pressure. Fluid results are normalized to the voltage amplitude applied in the PIC/MCC simulations for the respective conditions as indicated in table~\ref{tab:discharge}.
	\label{fig:PICvsFluid_V0}}%
\end{figure}

In order to evaluate the accuracy of the novel fluid model DDAn and the classical fluid model DDA53 for the considered discharge conditions, their results are compared to macroscopic quantities derived from the kinetic PIC/MCC simulations.
For this purpose, the amplitude of the current density $J_0$ obtained by the PIC/MCC simulations as indicated in table~\ref{tab:plasmaparameters} is used as input and the amplitude $V_0$ of the rf voltage is automatically adapted during the calculations according to equation~(\ref{eq:adaptivevoltage}). This is in accord with the procedure of Turner \textit{et al.}~\cite{Turner-2013-ID3004} who also used a fixed discharge current for benchmarking.
Figure~\ref{fig:PICvsFluid_V0} 
visualizes the amplitude $V_0^\mathrm{fluid}$ determined by the fluid modelling approaches DDAn and DDA53 in relation to the amplitude $V_0^\mathrm{PIC}$ prescribed in the PIC/MCC simulations (cf. table~\ref{tab:discharge}). 
It is found that the rf voltage required by the fluid models to reach the same current density $J_0$ is generally 10 to 30\,\% larger than that of the PIC/MCC simulations. This holds for both the fluid models.
Only the deviation of the rf voltage required by DDA53 to sustain the prescribed current density in argon increases monotonically with increasing pressure to 40\,\% at 80\,Pa.

\begin{figure}[ht]\centering%
	\includegraphics[clip,width=0.47\linewidth]{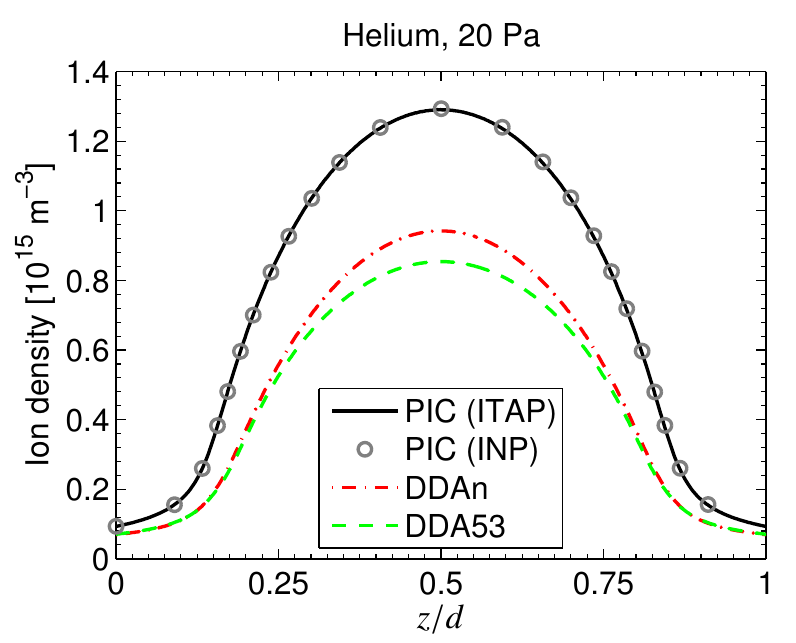}{\small~a)}\hfill%
	\includegraphics[clip,width=0.47\linewidth]{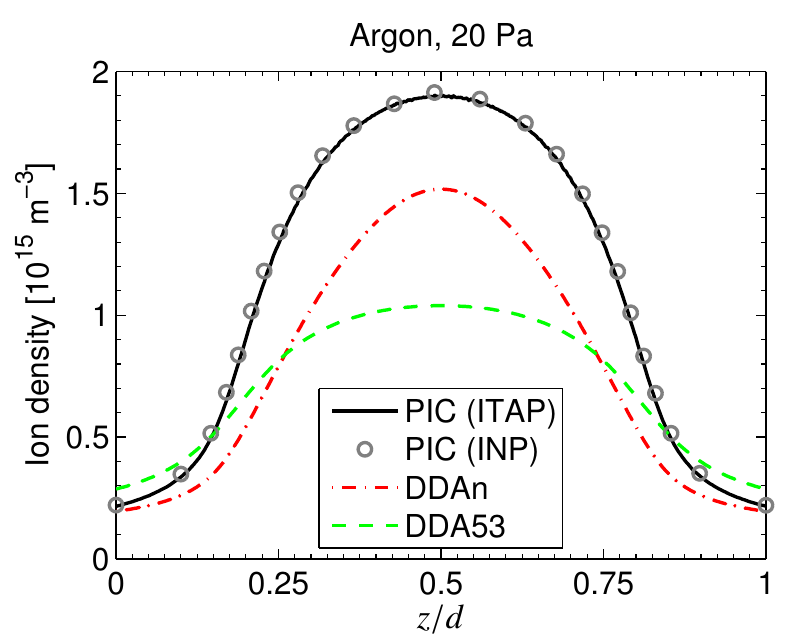}{\small~b)}%
	\caption{Spatial profile of the time averaged ion density $n_\mathrm{i}$ obtained by PIC/MCC simulations and the different fluid models in helium (a) and argon (b) at 20\,Pa.
	\label{fig:Vergleich_ni}}%
\end{figure}

Figure~\ref{fig:Vergleich_ni}
shows the results obtained by the two PIC/MCC simulation codes and the fluid models for the time averaged ion density at 20\,Pa in helium (figure~\ref{fig:Vergleich_ni}a) and argon (figure~\ref{fig:Vergleich_ni}b). 
As for the time and space averaged isotropic distribution $f_0/n_\mathrm{e}$ (see figure~\ref{fig:EEDF}), the predictions of the different PIC/MCC simulation codes PIC(ITAP) and PIC(INP) agree very well. Hence, the PIC/MCC results are not distinguished in the following. 
When comparing the fluid results with the PIC/MCC solution for helium in figure~\ref{fig:Vergleich_ni}a, it turns out that the spatial profile of the ion density predicted by the fluid models DDAn and DDA53 is in qualitative agreement with the PIC/MCC results. But both fluid model approaches underestimate the ion density by approximately 30\,\%. 
For argon (figure~\ref{fig:Vergleich_ni}b) the ion density obtained by DDAn is much closer to the PIC/MCC solution than that obtained by DDA53. The latter underestimates the ion density in the centre of the gap by almost 50\,\% but predicts larger values than PIC/MCC at the boundaries.

\begin{figure}[ht]\centering%
	\includegraphics[clip,width=0.47\linewidth]{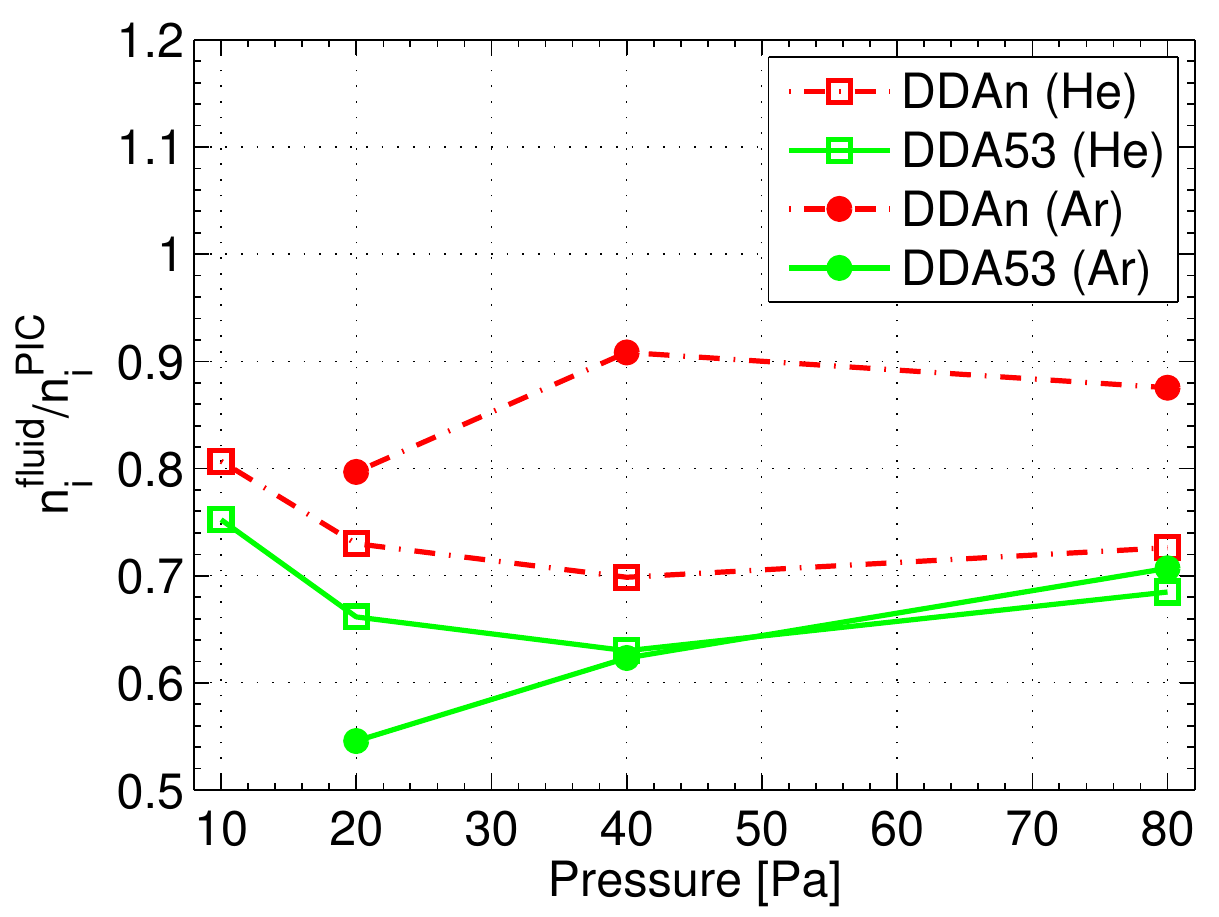}
	\caption{Time averaged ion density in the centre of the gap obtained by DDAn and DDA53 in helium and argon as a function of gas pressure. Fluid results are normalized to the PIC/MCC solution for the respective conditions as indicated in table~\ref{tab:plasmaparameters}.
	\label{fig:PICvsFluid_ni}}%
\end{figure}

Figure~\ref{fig:PICvsFluid_ni}
shows the maximum ion densities $n_\mathrm{i}^\mathrm{fluid}$ obtained by fluid modelling in relation to that of the PIC/MCC simulations $n_\mathrm{i}^\mathrm{PIC}$. 
It is found that the large deviation of the maximum ion density obtained by DDA53 at 20\,Pa in argon reduces to 30\,\% at a pressure of 80\,Pa. 
At the same time, the differences between DDAn and PIC/MCC for argon reduce from 20 to 10\,\% if the pressure is increased from 20 to 80\,Pa.
By contrast, the smallest deviation between PIC/MCC and both fluid modelling results for the maximum of the ion density in helium is obtained at the lowest pressure of 10\,Pa. This might be explained by the fact that the EVDF in helium at 10\,Pa is almost Maxwellian (cf. figure~\ref{fig:EEDF}a). In such situations fluid approaches are generally more adequate.

\begin{figure}[ht]\centering%
	\includegraphics[clip,width=0.47\linewidth]{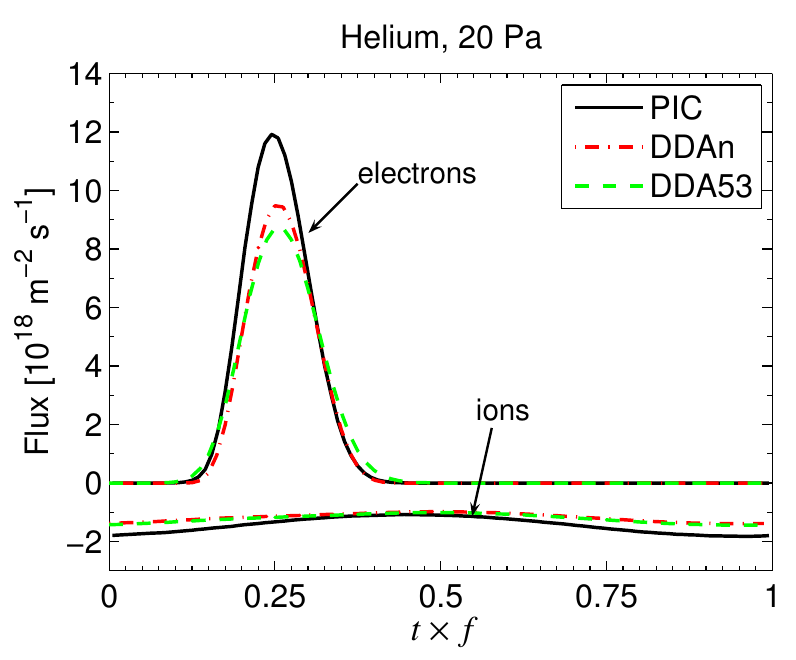}{\small~a)}%
	\includegraphics[clip,width=0.47\linewidth]{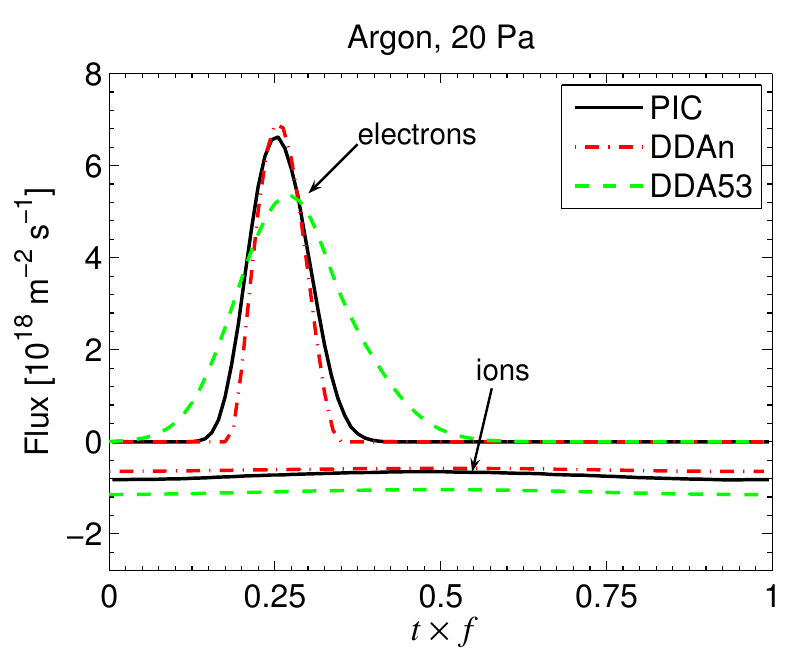}{\small~b)}%
	\caption{Temporal variation of the electron and ion fluxes $\mathit{\Gamma}_\mathrm{e,i}$ obtained by PIC/MCC simulations and the different fluid models at the powered electrode ($z=0$) in helium (a) and argon (b) at 20\,Pa. %
	\label{fig:Vergleich_JeJi}}%
\end{figure}

The temporal variation of the electron and ion fluxes at the powered electrode ($z=0$) are presented in figure~\ref{fig:Vergleich_JeJi} 
for helium and argon at 20\,Pa. Due to the symmetry of the discharge configuration, the same behaviour can be observed at the grounded electrode ($z=d$) with a time shift of $t\times f=0.5$.
Again, the results of both fluid modelling approaches are in qualitative agreement with the PIC/MCC simulations for helium.
However, only the novel drift-diffusion approximation DDAn is in conformity with the PIC/MCC simulations for argon while larger deviations are obtained when using the DDA53 fluid model.

\begin{figure}[ht]\centering%
	\includegraphics[clip,width=0.47\linewidth]{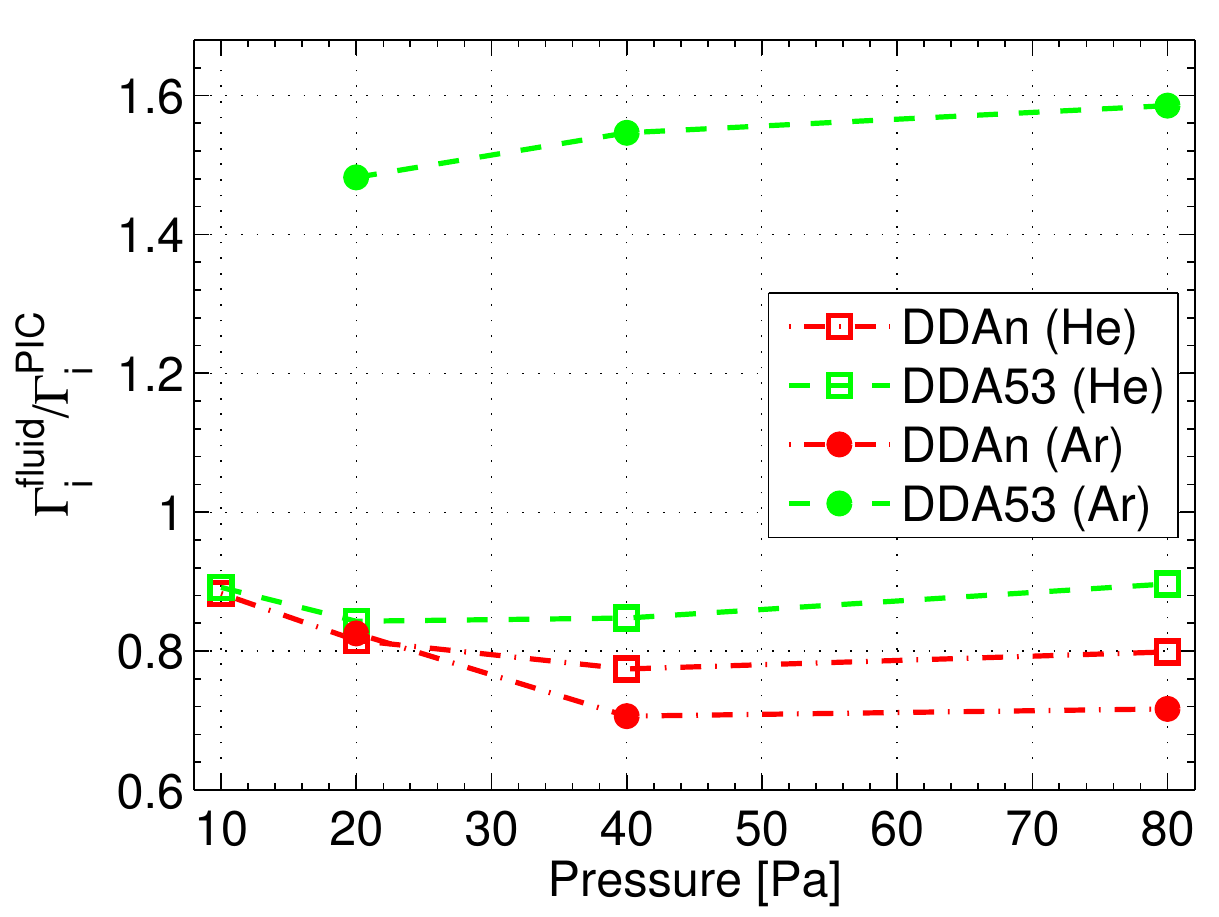}
	\caption{Time averaged ion flux at the powered electrode obtained by DDAn and DDA53 in helium and argon as a function of gas pressure. Fluid results are normalized to the PIC/MCC solution for the respective conditions as indicated in table~\ref{tab:plasmaparameters}.
	\label{fig:PICvsFluid_ji}}%
\end{figure}

The differences between the PIC/MCC simulation results and the fluid results for the particle fluxes at the electrodes are quantified in figure~\ref{fig:PICvsFluid_ji} by means of the time averaged ion fluxes $\mathit{\Gamma}_\mathrm{i}$ at the powered electrode.
Note that the accurate knowledge of this parameter is of particular importance for many applications~\cite{Donko-2012-ID3151}.
Figure~\ref{fig:PICvsFluid_ji} shows that for helium the fluid models DDA53 and DDAn underestimate the average ion flux by about 10 to 20\,\% compared to the PIC/MCC simulations. Here, the deviations for DDA53 are slightly smaller than for DDAn.
In argon, DDAn underestimates the average ion flux by about 20 to 30\,\%. By contrast, DDA53 largely overestimates the average ion flux in argon. The differences to the PIC/MCC simulation results increase from 50\,\% at 20\,Pa to 60\,\% at 80\,Pa. 

In summary, the plasma density and the average ion flux at the electrodes as ``global'' plasma parameters are predicted by the novel drift-diffusion approximation DDAn with an uncertainty of less than 30\,\% compared to PIC/MCC simulations. Much larger errors of up to 60\,\% are to be expected for these parameters for the classical fluid model DDA53 in argon. 
For helium, smaller deviations of less than 40\,\% are observed.
Similar differences between classical fluid models and PIC/MCC simulations have previously been reported for ccrf discharges in helium~\cite{Surendra-1995-ID3040} and argon~\cite{Kim-2005-ID2245} at comparable conditions.

\begin{figure}[ht]\centering%
	\includegraphics[clip,width=\linewidth]{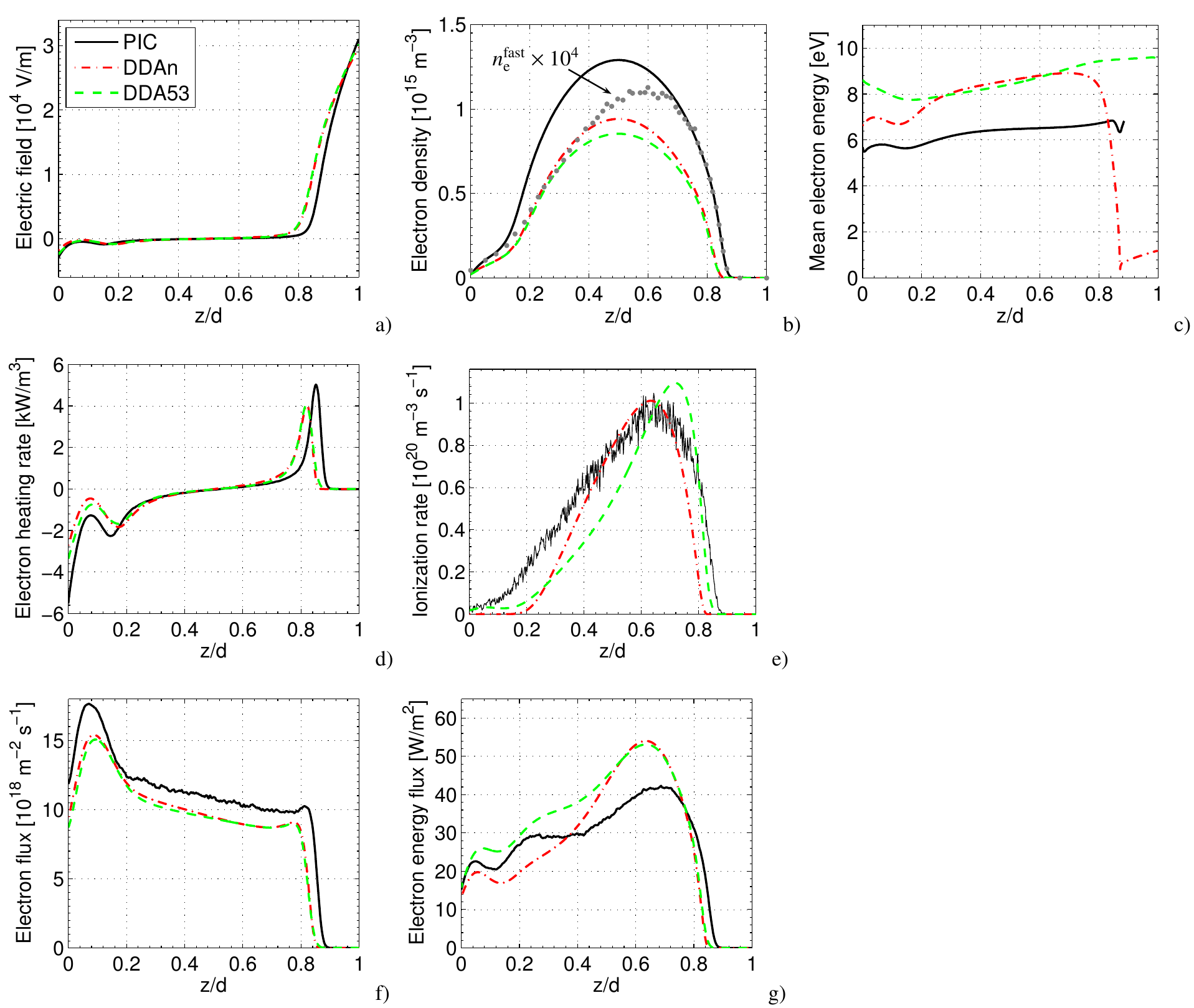}			
	\caption{Results for helium at 20\,Pa: Spatial variation of electric field $E$ (a), electron density $n_\mathrm{e}$ (b), mean electron energy $U_\mathrm{e}$ (c), electron heating rate $-e_0 E\mathit{\Gamma}_\mathrm{e}$ (d), ionization rate $S$ (e), particle flux $\mathit{\Gamma}_\mathrm{e}$ (f) and energy flux $Q_\mathrm{e}$ (g) of electrons obtained by the different modelling approaches at time $t\times f = 0.25$.%
	\label{fig:Vergleich0p25He}}%
\end{figure}

In order to get deeper insights into the differences of the considered fluid descriptions, the spatial variations of macroscopic quantities derived from fluid modelling and PIC/MCC simulations at the instant $t\times f = 0.25$ are compared in the following. 
At this time the voltage applied at $z=0$ reaches its maximum, $V_0$, and the momentary cathode is at $z=d$.
Figure~\ref{fig:Vergleich0p25He} 
shows the spatial variation of the first four moments of the EVDF, namely density, mean energy, particle flux and energy flux of electrons together with the electric field, the ionization rate and the electron heating rate, as obtained by the fluid models DDAn and DDA53 as well as the PIC/MCC simulations.
The density of fast electrons with energies higher than $e_0 V_0/4$ obtained by PIC/MCC simulations is additionally depicted in figure~\ref{fig:Vergleich0p25He}b.
In general, both fluid models are able to reproduce most discharge features qualitatively and quantitatively fairly well.
Larger differences can particularly be observed in the mean electron energy (figure~\ref{fig:Vergleich0p25He}c). Both fluid approaches overestimate the mean electron energy in the plasma bulk region by about 20--30\,\%. A large discrepancy can also be observed in the mean electron energy obtained by DDAn and DDA53 in the cathode region close to $z=d$. However, this is of minor importance for the present situation because no emission of secondary electrons is considered and hence the electron density in the sheath region is extremely low ($n_\mathrm{e}<\unit[1]{m^{-3}}$). Note that this is also the reason why no PIC/MCC data for the mean electron energy $U_\mathrm{e} = w_\mathrm{e}/n_\mathrm{e}$ is available for that region.
The overestimation of the mean electron energy in the bulk plasma by the fluid models is caused by the fact that, particularly at lower pressures, highly energetic electrons significantly contribute to the ionization budget. 
At the same time, the density of these fast electrons (figure~\ref{fig:Vergleich0p25He}b) is too low to have any effect on the mean electron energy. Since the ionization rate coefficient used in the fluid description depends on the mean electron energy only, (see section~\ref{sec:inp_phys}), a higher mean energy is enforced by the need to deliver the predefined current. 
This lack of a precise description of electrons contributing to the electron generation but not to the mean electron energy also causes a more localized profile of the ionization rate determined by the fluid models compared to the PIC/MCC results (figure~\ref{fig:Vergleich0p25He}e). 

Another important difference between the fluid modelling results and the PIC/MCC solution concerns the obtained sheath width. Particularly the spatial profiles of the electric field (figure~\ref{fig:Vergleich0p25He}a) and the fluxes (figures~\ref{fig:Vergleich0p25He}f,\,g) show that both fluid models overestimate the sheath width by about 5\,\%. 
This difference causes the spatial shift between the PIC/MCC and fluid results, e.g., for the maximum of the heating rate  in the sheath/plasma transition region (figure~\ref{fig:Vergleich0p25He}d).
Apart from this, the results of DDAn and DDA53 for the electron heating rate as well as the particle and energy fluxes are in good agreement with the PIC/MCC results. 
It is worth mentioning that the fluxes $\mathit{\Gamma}_\mathrm{e}$ and $Q_\mathrm{e}$ are highly transient quantities 
which makes the direct comparison of results obtained by different methods at a certain time  difficult.

\begin{figure}[ht]\centering%
	\includegraphics[clip,width=\linewidth]{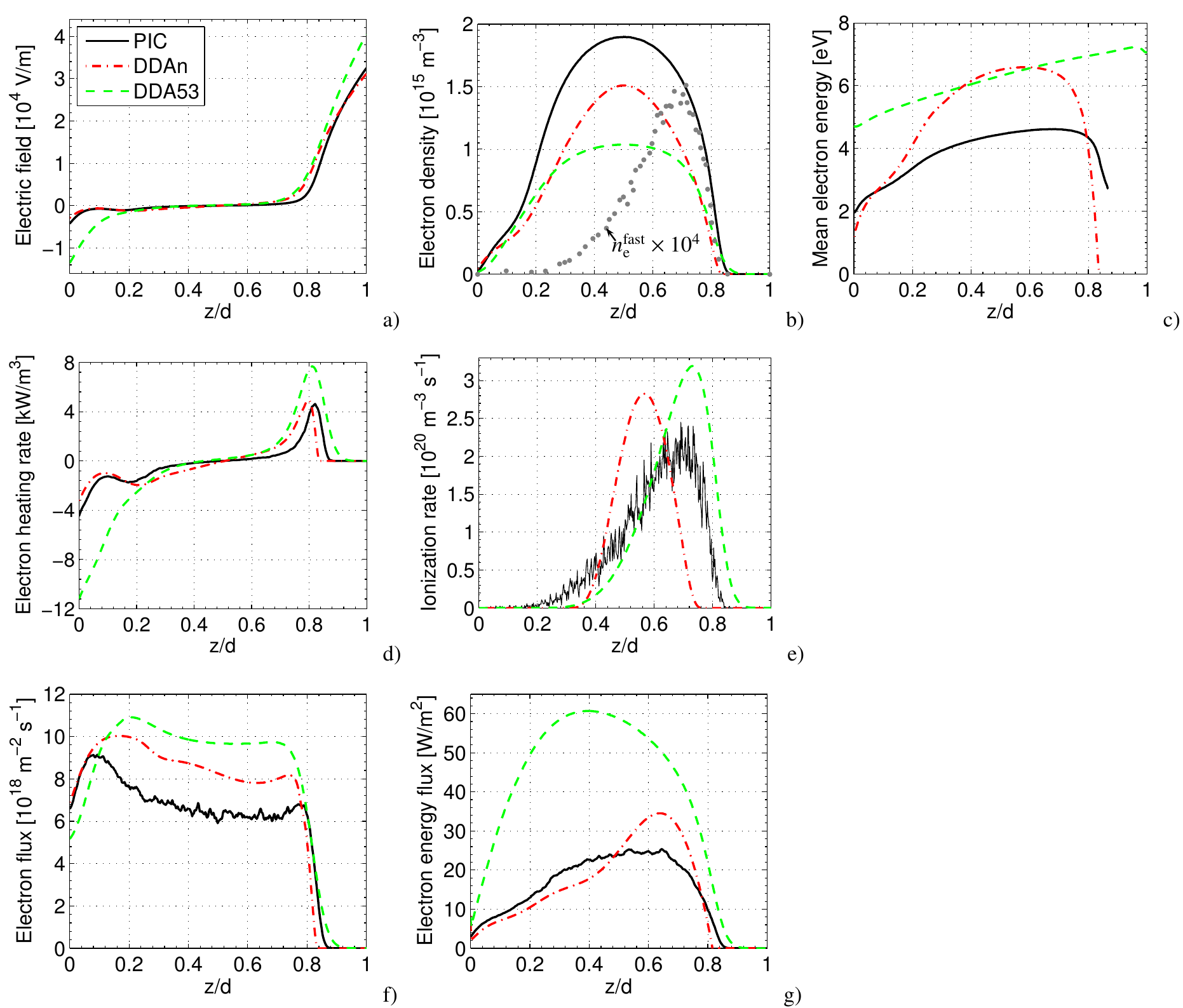}			
	\caption{Results for argon at 20\,Pa: Spatial variation of electric field $E$ (a), electron density $n_\mathrm{e}$ (b), mean electron energy $U_\mathrm{e}$ (c), electron heating rate $-e_0 E\mathit{\Gamma}_\mathrm{e}$ (d), ionization rate $S$ (e), particle flux $\mathit{\Gamma}_\mathrm{e}$ (f) and energy flux $Q_\mathrm{e}$ (g) of electrons obtained by the different modelling approaches at time $t\times f = 0.25$.%
	\label{fig:Vergleich0p25Ar}}%
\end{figure}

The corresponding spatial behaviour of the electric field, electron density, mean energy, heating and ionization rate, electron particle and energy fluxes for argon at 20\,Pa and the instant $t\times f=0.25$ is shown in figure~\ref{fig:Vergleich0p25Ar}.
The main difference to the situation for helium is that here only the novel fluid description for electrons, DDAn, is able to predict most of the macroscopic quantities with the same accuracy as in helium. Most likely, the presence of the Ramsauer minimum in argon causes the occurrence of nonlocal transport effects which can be captured by the DDAn approach but not by the classical model DDA53 due to the coinciding particle and energy transport of the electrons induced by the assumption of a Maxwellian EVDF~\cite{Becker-2013-ID2934}.
The divergences between the results of DDA53 and PIC/MCC particularly affect the electron heating rate (figure~\ref{fig:Vergleich0p25Ar}d) and the energy flux~(figure~\ref{fig:Vergleich0p25Ar}g) where large differences in the spatial profiles can be observed. In contrast, the fluid model DDAn including a consistent description of electron energy transport, provides a comparatively good prediction of these quantities when compared with the PIC/MCC simulation results. 
However, the comparison of the PIC/MCC results for the spatial profile of the fast electron density (figure~\ref{fig:Vergleich0p25Ar}b) with the spatial profile of the ionization rate (figure~\ref{fig:Vergleich0p25Ar}e) indicates that highly energetic electrons predominantly determine the ionization rate, similar to the behaviour found in helium. Hence, the marked differences between the results of DDAn and the PIC/MCC simulations for the spatial profiles of the mean electron energy  (figure~\ref{fig:Vergleich0p25Ar}c) and the ionization rate (figure~\ref{fig:Vergleich0p25Ar}e) 
are again caused by the improper consideration of ionization in the plasma bulk induced by fast electrons.


The approach introduced by Rafatov~\textit{et al.}~\cite{Rafatov-2012-ID2864} for low-pressure dc glow discharges aiming at an enhanced description of nonlocal ionization by adding an additional source term was found to be not applicable for the modelling of ccrf discharges. The separate description of highly energetic electrons by the Monte Carlo collision method (see, e.g.,~\cite{Bogaerts-1999-ID3030,Donko-1998-ID1246}) could be a more promising extension of the present fluid model description to overcome the remaining shortcomings.
Note that the discrepancy between fluid and PIC/MCC methods regarding the spatial distribution of the ionization rate drops with raising pressure due to the increase of the collisionality and the associated decrease of the impact of fast electrons.

\section{Conclusions}

In the present work the applicability and the accuracy of two different fluid approaches for the analysis of low-pressure ccrf discharges were investigated by benchmarking them against PIC/MCC simulations.
The considered fluid methods comprise time-dependent particle and momentum balance equations for ions as well as a novel drift-diffusion approximation (DDAn) and the classical drift-diffusion approximation with simplified energy transport coefficients (DDA53) for electrons, respectively.
In order to assure the general validity of the findings and to provide a test bed for future studies, simple ccrf discharge configurations in helium and argon at pressures ranging from 10 to 80\,Pa were considered.
Main findings of the comparative studies are the following:
\begin{itemize}
	\item Results of the novel as well as the classical fluid model are in good qualitative and quantitative agreement with macroscopic quantities derived from PIC/MCC simulations for ccrf discharges in helium.
	Here, the novel drift-diffusion approximation provides a slightly better prediction of the plasma density than the classical drift-diffusion approximation with deviations of less than 30\,\% for the DDAn model and 40\,\% for the DDA53 approach compared to PIC/MCC results.
	\item For argon, the classical fluid model fails to reproduce the discharge features predicted by PIC/MCC simulations. In contrast, the novel drift-diffusion approximation maintains its applicability and reliability and provides a prediction of relevant plasma parameters with deviations of less than 30\,\% compared to the PIC/MCC simulation results.
	\item At lower pressures both fluid models fail to correctly reproduce the spatial profile of the ionization rate. This is caused by the lack of an adequate description of highly energetic electrons which contribute to the ionization budget but not to the mean energy of electrons.
\end{itemize}
%
Future studies will address the questions how the influence of highly energetic electrons can be better included in the novel fluid modelling framework presented here and how plasma-boundary interactions can be described correctly.

\section*{Acknowledgment}

This work was partly supported by the German Research Foundation via SFB-TRR24 and by the PlasmaShape project from the European Union under grant agreement No 316216. The authors are grateful to A. Derzsi and Z. Donk\'o for many helpful discussions and their kindness to perform comparative benchmark calculations with their PIC/MCC code. 

\section*{References}


\providecommand{\newblock}{}



\end{document}